\documentclass[conference]{IEEEtran}
\IEEEoverridecommandlockouts

\usepackage{cite}
\usepackage{booktabs}
\usepackage[T1]{fontenc}
\usepackage{multirow}
\usepackage{amsmath,amssymb,amsfonts}
\usepackage{algorithmic}
\usepackage{graphicx}
\usepackage{textcomp}
\usepackage{xcolor}
\usepackage{hyperref}
\newcommand{\up}[1]{#1~\textcolor{green!60!black}{\footnotesize\ensuremath{\uparrow}}}
\newcommand{\down}[1]{#1~\textcolor{red}{\footnotesize\ensuremath{\downarrow}}}

\usepackage{url}
\def\BibTeX{{\rm B\kern-.05em{\sc i\kern-.025em b}\kern-.08em
    T\kern-.1667em\lower.7ex\hbox{E}\kern-.125emX}}

\usepackage{macros-review}
\usepackage{csquotes}
\usepackage{comment}

\usepackage[acronym,nomain]{glossaries}
\newacronym{os}{OS}{Operating System}
\newacronym{nyi}{NYI}{Not Yet Implemented}
\newacronym{dei}{DEI}{Department of Informatics Engineering}
\newacronym{llms}{LLMs}{Large Language Models}
\newacronym{llm}{LLM}{Large Language Model}
\newacronym{sft}{SFT}{Supervised Fine-tuning}
\newacronym{rlhf}{RLHF}{Reinforcement Learning with Human Feedback}
\newacronym{dpo}{DPO}{Direct Preference Optimization}
\newacronym{nlp}{NLP}{Natural Language Processing}
\newacronym{ai}{AI}{Artificial Intelligence}
\newacronym{slms}{SLMs}{Small Language Models}
\newacronym{ner}{NER}{Named Entity Recognition}
\newacronym{gdpr}{GDPR}{General Data Protection Regulation}
\newacronym{cwe}{CWE}{Common Weakness Enumeration}
\newacronym{samm}{SAMM}{Software Assurance Maturity Model}
\newacronym{xss}{XSS}{Cross-Site Scripting}
\newacronym{sqli}{SQLi}{SQL Injection}
\newacronym{waf}{WAF}{Web Application Firewall}
\newacronym{owasp}{OWASP}{Open Worldwide Application Security Project}
\newacronym{ipaas}{IPAAS}{Input PArameter Analysis System}
\newacronym{pda}{PDA}{Pushdown Automata}
\newacronym{lsp}{LSP}{Logic Score of Preference}
\newacronym{sat}{SAT}{Static Analysis Tool}
\newacronym{sast}{SAST}{Static Application Security Testing}
\newacronym{dast}{DAST}{Dynamic Application Security Testing}
\newacronym{sdlc}{SDLC}{Software Development Life Cycle}
\newacronym{iast}{IAST}{Interactive Application Security Testing}
\newacronym{rasp}{RASP}{Runtime Application Self-Protection}
\newacronym{qm}{QM}{Quality Model}
\newacronym{mcdm}{MCDM}{Multi-Criteria Decision-Making}
\newacronym{tpc}{TPC}{Transaction Processing Performance Council}
\newacronym{ssrf}{SSRF}{Server-Side Request Forgery}
\newacronym{sca}{SCA}{Software Composition Analysis}
\newacronym{bec}{BEC}{Business Email Commitments}
\newacronym{dtos}{DTOs}{Data Transfer Objects}
\newacronym{issre}{ISSRE}{The International Symposium on Software Reliability Engineering}

\newacronym{fp}{FP}{False Positive}
\newacronym{fn}{FN}{False Negative}
\newacronym{tp}{TP}{True Positive}
\newacronym{tn}{TN}{True Negative}
\newacronym{fpr}{FPR}{False Positive Rate}
\newacronym{fnr}{FNR}{False Negative Rate}
\newacronym{tpr}{TPR}{True Positive Rate}
\newacronym{tnr}{TNR}{True Negative Rate}
\newacronym{ppv}{PPV}{Positive Predictive Value}
\newacronym{npv}{NPV}{Negative Predictive Value}
\newacronym{roc}{ROC}{Receiver Operator Characteristic}
\newacronym{ci/cd}{CI/CD}{Continuous Integration and Continuous Deployment}
\newacronym{cot}{CoT}{Chain-of-Thought}
\newacronym{mae}{MAE}{Mean Absolute Error}
\newacronym{cert}{CERT}{Computer Emergency Readiness Team}
\newacronym{sei}{SEI}{Software Engineering Institute}

\newacronym{loc}{LoC}{Line of Code}

\newacronym{rq}{RQ}{Research Question}

\usepackage{eso-pic}

\begin{document}
\title{Leveraging Large Language Models for Trustworthiness Assessment of Web Applications}

\author{\IEEEauthorblockN{1\textsuperscript{st} Oleksandr Yarotskyi}
\IEEEauthorblockA{
\textit{University of Coimbra}\\
\textit{CISUC/LASI, DEI} \\
Coimbra, Portugal \\
oleksandr@student.dei.uc.pt}
\and
\IEEEauthorblockN{2\textsuperscript{nd} José D'Abruzzo Pereira}
\IEEEauthorblockA{
\textit{University of Coimbra}\\
\textit{CISUC/LASI, DEI} \\
Coimbra, Portugal\\
josep@dei.uc.pt}
\and
\IEEEauthorblockN{3\textsuperscript{rd} João R. Campos}
\IEEEauthorblockA{
\textit{University of Coimbra}\\
\textit{CISUC/LASI, DEI} \\
Coimbra, Portugal \\
jrcampos@dei.uc.pt}
}

\AddToShipoutPictureFG*{%
  \AtPageUpperLeft{%
    \raisebox{-1.0cm}{%
      \makebox[\paperwidth][c]{%
        \parbox{0.9\paperwidth}{%
          \centering
          \small\itshape
          Preprint version of a manuscript accepted for publication in the
          \textit{19th IEEE International Conference on Software Testing, Verification and Validation (ICST) 2026, Daejeon, Republic of Korea}.
        }%
      }%
    }%
  }%
}
\maketitle
\maketitle

\begin{abstract}
    The widespread adoption of web applications has made their security a critical concern and has increased the need for systematic ways to assess whether they can be considered trustworthy. 
    However, \enquote{trust} assessment remains an open problem as existing techniques primarily focus on detecting known vulnerabilities or depend on manual evaluation, which limits their scalability; therefore, evaluating adherence to secure coding practices offers a complementary, pragmatic perspective by focusing on observable development behaviors.
    In practice, the identification and verification of secure coding practices are predominantly performed manually, relying on expert knowledge and code reviews, which is time-consuming, subjective, and difficult to scale. 
    This study presents an empirical methodology to automate the trustworthiness assessment of web applications by leveraging Large Language Models (LLMs) to verify adherence to secure coding practices. We conduct a comparative analysis of prompt engineering techniques across five state-of-the-art LLMs, ranging from baseline zero-shot classification to prompts enriched with semantic definitions, structural context derived from call graphs, and explicit instructional guidance. Furthermore, we propose an extension of a hierarchical Quality Model (QM) based on the Logic Score of Preference (LSP), in which LLM outputs are used to populate the model’s quality attributes and compute a holistic trustworthiness score. Experimental results indicate that excessive structural context can introduce noise, whereas rule-based instructional prompting improves assessment reliability. The resulting trustworthiness score allows discriminating between secure and vulnerable implementations, supporting the feasibility of using LLMs for scalable and context-aware trust assessment.
\end{abstract}



\section{Introduction}
\label{sec:intro}

Web applications are fundamental for the modern digital infrastructure, supporting critical operations across a range of sectors~\cite{antunesDefendingWebApplication2012, zhuSurveySecurityAnalysis2024}. As a consequence, we increasingly depend on these systems to behave correctly and securely, making their trustworthiness a critical quality attribute.
However, this makes them prime targets for cyber threats, often stemming from insecure coding practices~\cite{antunesDefendingWebApplication2012, faisalfadlallaInputValidationVulnerabilities2023}.
The persistence of these threats is exemplified by the massive 2023 \textit{MOVEit} data breach, where a single SQL Injection flaw (CVE-2023-34362) compromised thousands of organizations globally~\cite{lakshmananNewCriticalMOVEit2023}.
With injection flaws consistently ranking among the top risks in the \gls{owasp} Top 10~\cite{owaspTop10}, it is evident that fundamental practices remain a critical yet often neglected line of defense~\cite{mitropoulosDefendingWebApplication2019,ibrahimkhalafWebAttackDetection2021,sharPredictingCommonWeb2012,scholtePreventingInputValidation2012}.

Traditional security assessment approaches focus primarily on detecting known vulnerabilities~\cite{antunesDefendingWebApplication2012, faisalfadlallaInputValidationVulnerabilities2023}. However, this perspective is limited: the absence of alerts does not guarantee the reliability of the application, as it ignores latent weaknesses~\cite{Pereira2023}. As a result, vulnerability-centric security tools are insufficient to support a comprehensive trustworthiness assessment. Additionally, conventional techniques have structural limitations: \gls{sast} tends to generate high \glspl{fp} due to a lack of context, while \gls{dast} fails to cover critical execution paths~\cite{DAbruzzoPereira2020, singhAnalysisWebApplication2024, zhuComprehensiveStudyStatic2024}. Hybrid solutions (\gls{iast}, \gls{rasp}), while attempting to mitigate these failures, often introduce prohibitive complexity and execution overhead~\cite{owasp2024, snykiast2024}.


To overcome these limitations, \textit{Lemes et al.}~\cite{lemesTrustworthinessAssessmentWeb2019} proposed a more proactive approach that focuses on assessing trustworthiness as the primary metric for measuring security quality. Rather than solely identifying individual flaws, this approach evaluates software by assessing its adherence to established secure coding practices and security-related quality attributes. The central hypothesis is that applications that demonstrate higher adherence to proven secure coding practices (such as those defined by \gls{owasp}~\cite{owaspSecureCodingPractices2025}) exhibit a lower likelihood of containing unknown vulnerabilities and can therefore be considered more trustworthy~\cite{lemesTrustworthinessAssessmentWeb2019}. However, approaches for assessing trustworthiness based on coding practices, such as that proposed by~\cite{lemesTrustworthinessAssessmentWeb2019} and recently extended to characterize open-source software security~\cite{Pereira2023}, remain challenging to implement in practice. While robust, these approaches rely on manual or semi-automated verification to determine adherence to secure coding practices, limiting their scalability and integration into agile development environments, where automation is essential.

At the same time, \glspl{llm} have recently demonstrated remarkable capabilities in software engineering~\cite{Vieira2025}, including understanding and generating code~\cite{xuSystematicEvaluationLarge2022, zhaoSurveyLargeLanguage2024}, opening new perspectives for cybersecurity~\cite{bezziLargeLanguageModels2024}.
Emerging applications include direct vulnerability detection~\cite{fuChatGPTVulnerabilityDetection2023, khareUnderstandingEffectivenessLarge2024, tambergHarnessingLargeLanguage2024}, automatic code repair \cite{islamLLMPoweredCodeVulnerability2024, zhouMultiLLMCollaborationDataCentric2024}, and providing advice on secure coding \cite{sloane-andersonBenchmarkingLargeLanguage2024, espinhagasibaImSorryDave2023}. These capabilities suggest that \glspl{llm} may be well-suited to assess coding practices that require semantic understanding and contextual reasoning beyond pattern-based analysis. 

In this work, we investigate the use of \glspl{llm} as a systematic mechanism for evaluating source code against established secure coding standards for trustworthiness assessment. By moving beyond ad hoc or advisory uses of \glspl{llm}, this study explores their role in structured, repeatable security evaluation processes, a direction that remains insufficiently characterized in existing research~\cite{heLargeLanguageModels2023}. Building on the trustworthiness assessment framework proposed by \textit{Lemes et al.}~\cite{lemesTrustworthinessAssessmentWeb2019}, this study is guided by the following \glspl{rq}:

\begin{itemize}
    \item \textbf{\gls{rq}1:} Can \glspl{llm} identify adherence to \gls{owasp} Input Validation secure coding practices under baseline (zero-shot) prompting conditions?
    
    \item \textbf{\gls{rq}2:} What is the impact of different prompt context enrichment strategies on the accuracy and consistency of \gls{llm}-based secure coding practice identification?
    
    \item \textbf{\gls{rq}3:} To what extent can \glspl{llm} be used to produce reliable trustworthiness scores?
\end{itemize}

In summary, the main contributions of this paper are:

\begin{itemize}
    \item A systematic \textbf{comparison of prompt engineering strategies} for secure coding practice assessment, evaluating the effects of semantic enrichment, structural context, and explicit instructions across multiple \glspl{llm}.

    \item An empirical assessment of the \textbf{feasibility and reliability} of \gls{llm}-driven trustworthiness automation, clarifying when \gls{llm}-based analysis can be applied at scale.

    \item \textbf{An extension of the \gls{qm}} proposed by \textit{Lemes et al.}~\cite{lemesTrustworthinessAssessmentWeb2019}, leveraging \gls{llm} outputs to automate the classification of functions and compute trustworthiness scores.
\end{itemize}

This paper is organized as follows. Section~\ref{sec:back&rel} presents the background and related work. Section~\ref{sec:methodology} describes the experimental methodology. Section~\ref{sec:results_discussion} reports the evaluation results and the discussion. Section~\ref{sec:threats} addresses threats to validity, and Section~\ref{sec:conclusion} concludes the paper and outlines directions for future work.


\section{Background and Related Work}
\label{sec:back&rel}
This section presents background and related work of secure coding practices, trustworthiness assessment, and \glspl{llm}.

\subsection{Secure Coding Practices}
\label{sec:secure_coding}

A fundamental strategy to enhance the security of web applications is to implement security measures in the early stages of the development cycle~\cite{zhuSurveySecurityAnalysis2024}.
To guide this process, diverse frameworks have been established, derived from industry experience. Initiatives like the Building Security in Maturity Model (BSIMM)~\cite{bsimm2018} focus on measuring organizational maturity across various domains. In contrast, the Microsoft Security Development Lifecycle (SDL)~\cite{microsoftSDL} integrates security and privacy throughout the entire development lifecycle.

Regarding specific code-level rules, the \gls{cert} division from the \gls{sei} provides distinct lists of secure coding practices for languages such as C, C++, and Java~\cite{seiCertCoding}. These practices aim to eliminate insecure behaviors that frequently lead to vulnerabilities. 
However, \gls{cert} rules often lack web-specific context (\textit{e.g.}, HTTP request handling) and are too fragmented across languages to serve as a unified standard for web trustworthiness~\cite{seiCertTop10}.


The \gls{owasp} Secure Coding Practices~\cite{owaspSecureCodingPractices2025} constitute a comprehensive reference framework for the development of secure web applications. They provide a structured set of implementation-level guidelines for secure web application development. The practices span multiple security-relevant dimensions, including controlled data ingestion and processing, authentication and session state management, authorization enforcement, robust error handling and logging, cryptographic key and data management, and secure interaction with external services and dependencies~\cite{owasp}. 

For implementation to be effective, secure coding practices must be applied within a defense-in-depth strategy, incorporating server-side enforcement and complementary safeguards. However, despite the availability of comprehensive guidelines, several studies highlight a persistent \textit{gap in implementation}, as developers often apply secure coding practices inconsistently or incompletely due to time and resource constraints~\cite{scholtePreventingInputValidation2012, faisalfadlallaInputValidationVulnerabilities2023}.


\subsection{Trustworthiness Assessment}
\label{sec:trust_assessment}
Evaluating the security of software requires moving beyond simple vulnerability counting toward holistic, quantitative metrics. To address this challenge, the community has increasingly adopted \gls{mcdm} methods to aggregate diverse security criteria~\cite{mcdm, jiangTrustworthinessEvaluationMethod2014, macekModelEvaluationCritical2021, ningHybridMCDMApproach2020}.

The \gls{lsp} methodology is a technique of the \gls{mcdm} and is a quantitative quality assessment framework based on logical aggregation operators that simultaneously explicitly model the logical relationships between criteria~\cite{dujmovicModelingAggregationSecurity2012}. This methodology is distinguished by its superior ability to model complex logical requirements, specifically \textit{simultaneity}, where distinct criteria must be satisfied together, and \textit{replaceability}, where a strong attribute can compensate for the absence of another~\cite{dujmovicModelingAggregationSecurity2012}.

Its specific applicability to security standards has been validated by \textit{Dasso and Funes}~\cite{dassoWebApplicationsSecurity2020}, who successfully employed \gls{lsp} to assess compliance with the \gls{owasp} Web Security Testing Guide.
By modeling the guide testing controls into a logic-based hierarchical structure, they demonstrated that \gls{lsp} generates a global security indicator that effectively distinguishes between minor and critical non-conformity.
Their results validated \gls{lsp} as a robust framework for scoring security compliance, surpassing simple additive models, which often fail to capture the severity of missing critical controls~\cite{dassoWebApplicationsSecurity2020}.

Building on this foundation, \textit{Lemes et al.}~\cite{lemesTrustworthinessAssessmentWeb2019} applied \gls{lsp} to the granular level of secure coding. 
They proposed a proactive framework using a \gls{qm} to hypothesize that applications implementing secure practices have a lower risk of unknown vulnerabilities.
The framework employs a hierarchical structure where leaf nodes represent individual \gls{owasp} input validation practices, intermediate nodes represent categories, and the root represents overall trustworthiness~\cite{lemesTrustworthinessAssessmentWeb2019}.
A key innovation of this framework is the derivation of weights from real-world vulnerability data. Practices linked to frequent vulnerabilities receive higher importance coefficients, ensuring that the final scalar score reflects empirical risk rather than subjective judgment~\cite{lemesTrustworthinessAssessmentWeb2019}.

\textit{Pereira and Vieira}~\cite{Pereira2023} extended the \gls{lsp}-based characterization to the domain of open-source software. 
Their work focused on characterizing the security of C/C++ functions by aggregating software metrics, demonstrating the flexibility of \glspl{qm} to evaluate security at a function level beyond web applications.
Their results demonstrated that aggregating static analysis metrics via \gls{lsp} provided an accurate ranking of critical functions from a security perspective.
A validation is performed with security experts, and their automated approach can be used to reduce the time spent to do security assessment, as experts are not involved in such assessment.

Despite their relevance, these approaches rely on manual code inspection or rule-based static analysis, limiting their scalability in modern \gls{ci/cd} pipelines.

\subsection{Large Language Models}
\label{sec:llms_security}
\glspl{llm} are types of neural networks that are trained on vast corpora of natural language as and source code, which demonstrate strong capabilities for both pattern and semantic analysis as well as tracking dependencies across procedural boundaries~\cite{zhaoSurveyLargeLanguage2024, xuSystematicEvaluationLarge2022, heLargeLanguageModels2023}. 
These traits position \glspl{llm} as promising candidates for automated security assessment~\cite{bezziLargeLanguageModels2024}. However, their stochastic nature can produce plausible but incorrect assertions (hallucinations), necessitating rigorous empirical benchmarking before deployment in critical pipelines~\cite{weiChainofThoughtPromptingElicits2022}.

Existing applications of \glspl{llm} to software security can be broadly categorized into three main areas:
(1) \textbf{Vulnerability Detection}, which focuses on identifying security flaws in software artifacts by analyzing source code at different levels of granularity, including entire files or modules, individual functions or methods, and localized code fragments or snippets. This often relies on benchmark datasets such as Juliet~\cite{juliet-test-suite}, but also explores alternative corpora and hybrid approaches that combine \glspl{llm} with static or semantic code analysis techniques~\cite{fuChatGPTVulnerabilityDetection2023, khareUnderstandingEffectivenessLarge2024, tambergHarnessingLargeLanguage2024};
(2) \textbf{Automated Repair}, which leverages \glspl{llm} to generate or refine security patches for identified flaws, using approaches such as reinforcement learning~\cite{islamLLMPoweredCodeVulnerability2024}, multi-agent systems~\cite{zhouMultiLLMCollaborationDataCentric2024}, or feedback-driven code synthesis; and
(3) \textbf{Secure Coding Guidance}, where \glspl{llm} are employed as conversational assistants to provide contextual explanations, recommendations, and best-practice guidance aimed at preventing vulnerabilities and supporting secure software development~\cite{sloane-andersonBenchmarkingLargeLanguage2024, espinhagasibaImSorryDave2023}.



\section{Methodology}
\label{sec:methodology}
\begin{figure*}[htbp]
    \centering
    \includegraphics[width=1\textwidth]{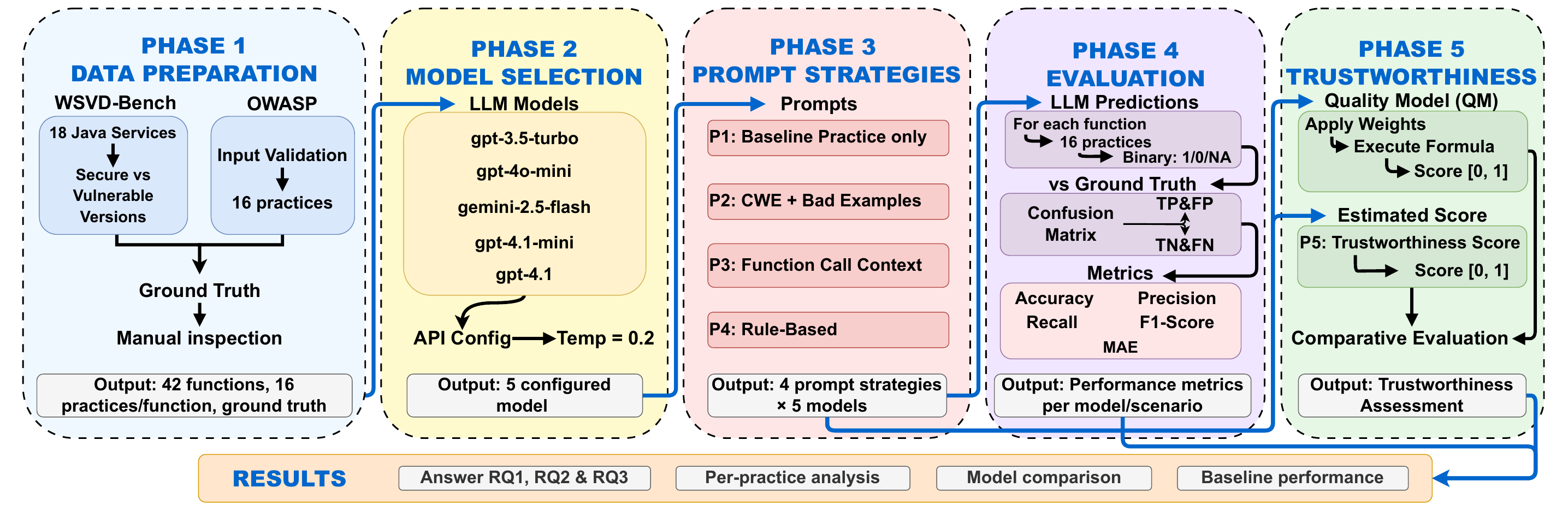}
    \caption{Overview of the methodology structure}
    \label{fig:struc-meth}
\end{figure*}

Manual evaluation of secure coding practices is time-consuming, expertise-intensive, and difficult to scale to modern software systems. To address this limitation, we investigate the use of \glspl{llm} to automate trustworthiness assessment of web applications through a systematic five-step methodology, illustrated in Fig.~\ref{fig:struc-meth}, designed to answer the three previously defined research questions. The methodology is organized into three stages: (i) prompt-based inference, where \glspl{llm} generate practice-level assessments or trustworthiness score estimates; (ii) definition of evaluation mechanisms, including manually curated ground truth and performance metrics; and (iii) trustworthiness assessment, aggregating results into system-level scores using a \gls{qm} and the \gls{lsp} methodology.


Due to time and space constraints, we focus on the \gls{owasp} Input Validation domain as a representative and security-critical subset, following a similar scoping strategy to \textit{Lemes et al.}~\cite{lemesTrustworthinessAssessmentWeb2019}. Input Validation consistently ranks among the most prevalent and impactful categories in the \gls{owasp} Top 10 (including the 2025 edition) and remains a primary source of injection vulnerabilities in modern web applications. This focused scope enables a controlled and in-depth evaluation while preserving the generality of the proposed methodology, which can be extended to other secure coding domains.


\subsection{Dataset}
\label{sec:dataset}
For this study, and to facilitate comparison with previous works, we use the WSVD-Bench Dataset \cite{wsvd-bench}, a public benchmark for web service vulnerabilities also employed by Lemes et al. \cite{lemesTrustworthinessAssessmentWeb2019}, enabling direct comparison with their manual assessment approach.

WSVD-Bench consists of Java web services derived from the TPC-App, TPC-C, and TPC-W benchmarks. The dataset comprises 21 distinct services and 80 service operations, including a total of 158 documented SQL Injection vulnerabilities~\cite{wsvd-bench}. Each service is provided in multiple variants: secure implementations without known vulnerabilities (\textit{Vx0}), fully vulnerable implementations (\textit{VxA}), and single-vulnerability variants targeting specific code locations (\textit{VxLine})~\cite{wsvd-bench, lemesTrustworthinessAssessmentWeb2019}.

This Vx0 vs. VxA duality provides clear examples of code adhering to (Vx0) and violating (VxA) \gls{owasp} secure coding practices \cite{owaspSecureCodingPractices2025}, specifically for SQL Injection prevention, which is consistently identified as a critical risk in web applications and has appeared in multiple editions of the OWASP Top~10~\cite{owaspSQLInjectionPrevention2024, owaspTop10}.

\subsection{OWASP Input Validation Practices}
\label{sec:input}

We evaluate adherence to the 16 \gls{owasp} Input Validation practices defined in the Secure Coding Practices Quick Reference Guide \cite{owaspSecureCodingPractices2025}. 
These practices address critical defenses such as server-side enforcement (Practice \#1), centralized validation routines (Practice \#3), validation of all client-provided data (Practice \#8), and neutralization of dangerous characters (Practice \#15), which collectively aim to ensure robust handling of untrusted input and reduce the risk of a wide range of input-driven vulnerabilities.

For each function and practice, we assess two dimensions: (1) \textbf{Applicability}: whether the practice is relevant to the function input handling logic; and (2) \textbf{Adherence}: if applicable, whether the practice is correctly implemented.

\subsection{Model Selection}
\label{sec:models}
We selected five widely used \glspl{llm} from multiple vendors, accessed via API with deterministic settings (temperature = 0.2) to ensure reproducibility. This diverse selection spans a range of capacities, enabling the analysis of performance trade-offs between efficiency-oriented and high-capacity models currently accessible to DevSecOps teams. Furthermore, all selected models have demonstrated strong code understanding capabilities in prior evaluations~\cite{xuSystematicEvaluationLarge2022, zhaoSurveyLargeLanguage2024}.

The selected models include: (1) \textbf{gpt-3.5-turbo}, serving as a cost-effective baseline; (2) \textbf{gpt-4o-mini}, \textbf{gpt-4.1-mini}, and Google \textbf{gemini-2.5-flash}, representing models optimized for low latency and efficient reasoning; and (3) \textbf{gpt-4.1}, reflecting a high-capacity model designed for complex reasoning tasks.

Regarding validity, although WSVD-Bench is public~\cite{antunesDefendingWebApplication2012} and likely included in the model's pre-training, the ground truth for these specific practices was manually constructed for this study. Consequently, even if the models have memorized the code structure, they cannot rely on training data to retrieve the classification labels required by our methodology, compelling them to perform active reasoning rather than mere repetition.

\subsection{Prompts Strategies}
\label{sec:prompts}
We designed a set of prompt templates to elicit secure coding practice assessments and trustworthiness estimates from \glspl{llm} to address the research questions. 
The Prompt~1 addresses \textbf{\gls{rq}1} through zero-shot practice identification. The Prompts 2–4 support \textbf{\gls{rq}2} by analyzing the effect of additional context on assessment reliability. \textbf{\gls{rq}3} is addressed by aggregating \gls{llm}-derived practice assessments using the \gls{qm} and by directly eliciting trustworthiness scores from \glspl{llm} for comparison with the reference model.

\textbf{Prompt 1 (Baseline):}
The baseline prompt provides only the \gls{owasp} practice description \cite{owaspSecureCodingPractices2025} and the Java function, requiring binary responses (\texttt{Yes/No}) for applicability and adherence. 
This zero-shot approach establishes the lower bound of \gls{llm} performance when relying solely on pre-trained knowledge, serving as a reference for evaluating contextual enrichment strategies.

\textbf{Prompt 2 (CWE Identifiers + Bad Examples):}
This prompt enriches the baseline through semantic grounding using two mechanisms: (1) mapping each secure coding practice to relevant \gls{cwe} categories defined by MITRE~\cite{cwe} (e.g., Practice~\#1 $\rightarrow$ \gls{cwe}-20, \gls{cwe}-602, \gls{cwe}-807); and (2) providing short Java code snippets (3–5 lines) illustrating representative violations. This few-shot strategy is intended to help \glspl{llm} recognize insecure coding patterns, thereby reducing false negatives. To ensure experimental control, \gls{cwe} mappings and example snippets are held constant across all functions for a given practice, so observed performance differences reflect model behavior rather than prompt variability.

\textbf{Prompt 3 (Function Call Context):}
To address validation logic delegated to helper functions, we use Joern \cite{joernDocs} to extract code property graphs and retrieve the source code of all called functions. 
This structural context enables \glspl{llm} to trace security practices across inter-procedural dependencies. 
This approach prevents false negatives in modular code where main functions defer validation to helper routines (\textit{e.g.}, a \texttt{sanitize()} call). 
Without visibility into invoked functions, \glspl{llm} may incorrectly classify compliant functions.

\textbf{Prompt 4 (Rule-Based):}
Adopting a structured assessment approach, this prompt defines clear binary conditions for compliance, removing the need for open interpretation.
Unlike previous prompts, it provides strict decision logic.
For example, regarding \textit{Practice 1} (server-side validation), the instruction explicitly defines the states:

\begin{quote}
    \small
    \textbf{INSTRUCTIONS} \\
    These instructions specify the conditions under which the rule is considered followed or not followed.
    \begin{itemize}
        \item \textbf{Followed:} The method is executed on the server (e.g., Java backend) and performs server-side validation;
        \item \textbf{Not followed:} The method relies exclusively on client-side validation (e.g., assumes the input is safe because it was validated in JavaScript);
\end{itemize}
\end{quote}

By applying this structure, the prompt lets us check if clear instructions are enough to make things more dependable, without needing to use \gls{cwe} definitions or outside examples.

\subsection{Evaluation}
\label{sec:evaluation}
For systematic, quantitative analysis of \gls{llm} performance, we designed an evaluation framework integrating manually labeled ground truth, the \gls{qm} from \cite{lemesTrustworthinessAssessmentWeb2019}, and metrics capturing multiple performance dimensions for both classification (Prompts 1-4) and scoring (Trustworthiness Scoring Prompt).

\textbf{Ground Truth Construction:}
\label{sec:gt}
The foundation of our evaluation is a manually curated ground truth dataset constructed through expert analysis of all functions in the WSVD-Bench corpus \cite{wsvd-bench}. 
For each of the 42 Java functions and each of the 16 \gls{owasp} Input Validation practices experts performed manual code inspection to determine practice adherence.

A practice is assigned \textit{1 (Followed)} if applicable and properly implemented, \textit{0 (Not Followed)} if applicable but missing or improperly implemented, and \textit{NA (Not Applicable)} if irrelevant to the function context (\textit{e.g.}, input validation practices in pure utility functions without external input sources).

\textbf{Performance Metrics:}
\label{sec:metrics}
For the evaluation of secure coding practice classification (Prompts~1-4), we compute standard multi-class metrics: \textit{Accuracy}, \textit{Precision}, \textit{Recall}, and \textit{F1-score}. 
The F1-score is used as the primary comparison metric, as it balances precision and recall and is robust to class imbalance across the three target labels (1/0/NA).

All metrics are macro-averaged across classes, assigning equal weight to each label. This avoids bias toward the majority class and ensures comparability across practices with differing applicability frequencies.

\subsection{Trustworthiness}
\label{sec:qm} 
To assess reliability, simply counting vulnerabilities is insufficient; we need to quantify the degree to which a code adheres to security standards. This step addresses \textbf{RQ3} by evaluating the viability of generating system-level scores. To this end, we contrast two approaches: score generation by \gls{llm} and a formal calculation using a structured \gls{qm} based on the \gls{lsp} methodology proposed by \textit{Lemes et al.}~\cite{lemesTrustworthinessAssessmentWeb2019}.

\textbf{Trustworthiness Scoring Prompt:}
\label{sec:prompt5}
Beyond binary classification, we evaluate whether \glspl{llm} can implicitly aggregate practice assessments into holistic trustworthiness scores $\in [0,1]$. 
The prompt provides: (1) the target function, (2) per-practice results from Prompts 1-4, and (3) a request for a scalar score with justification.
This tests whether \glspl{llm} can reproduce \gls{lsp}-weighted aggregation (Section \ref{sec:qm}) without an explicit formula.
To quantify the accuracy of these holistic estimates, we calculate the \gls{mae} between the \gls{llm}-generated scores and the ground truth scores derived from the formal \gls{qm}.

\begin{figure}[htbp]
    \centering
    \includegraphics[width=1\columnwidth]{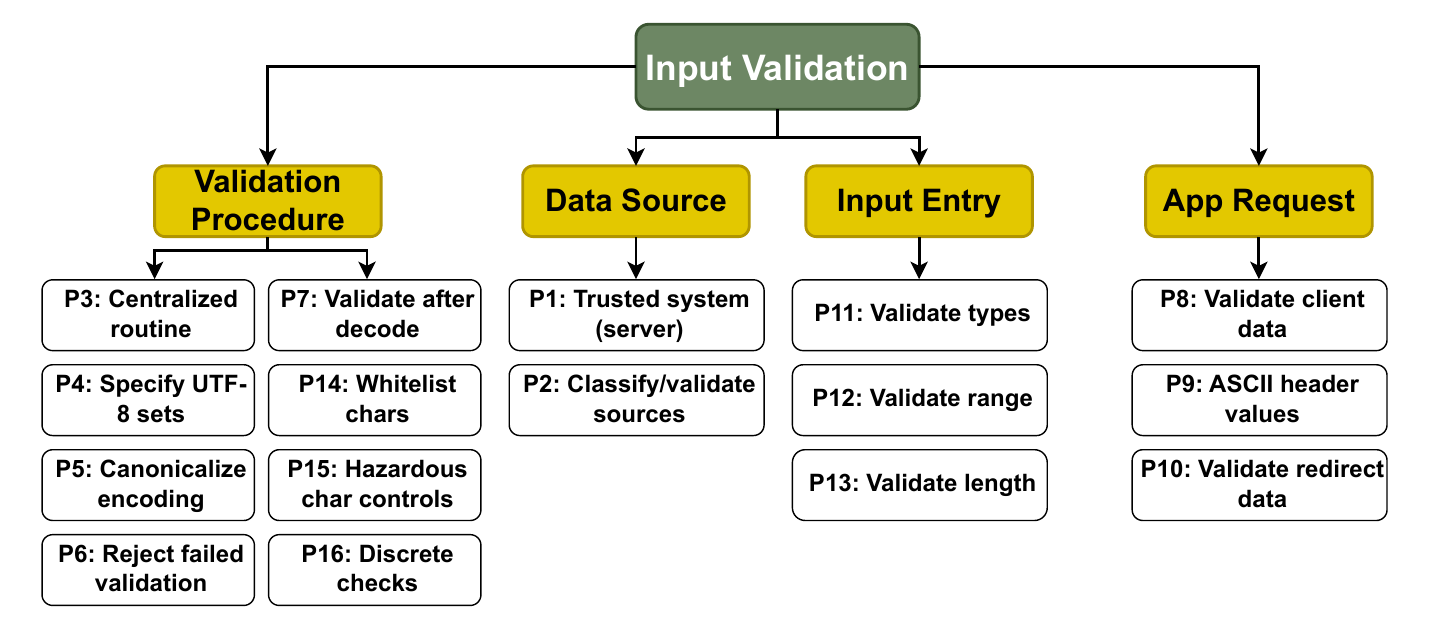}
    \caption{Quality Model Hierarchy of OWASP Input Validation Practices \cite{owaspSecureCodingPractices2025}. Adapted from Lemes et al. \cite{lemesTrustworthinessAssessmentWeb2019}.}
    \label{fig:qm_hierarchy}
\end{figure}

\textbf{Hierarchical Quality Model and LSP Aggregation:}
\label{sec:qm-t-def}
The \gls{qm} acts as a structured map of security requirements. It breaks down the abstract concept of ``Trustworthiness'' into smaller, measurable components. Specifically, it relies on the \gls{lsp} methodology to aggregate attributes into a global preference score.
Formally, the model is defined as a tuple $QM = (A, T, W, O)$, where $A$ represents the set of attributes (secure coding practices), $T$ the normalization thresholds, $W$ the relative weights, and $O$ the aggregation operators~\cite{lemesTrustworthinessAssessmentWeb2019}.
It organizes \gls{owasp} Input Validation practices hierarchically (see Figure~\ref{fig:qm_hierarchy}), where each applicable practice is encoded as a binary value $r_i \in \{0,1\}$ (1 = followed, 0 = not followed). Practices labeled as NA are excluded, with their weight distributed among applicable practices to maintain normalization.

The challenge lies in aggregating these individual checks. A standard weighted average is often inadequate for security, as it allows a strong attribute to compensate for a weak one (a property known as replaceability). To avoid this, the \gls{lsp} methodology uses logical aggregation operators. Specifically, we employ the property of simultaneity, which, in this context, implies that for an application to be reliable, it must satisfy all critical security criteria simultaneously. Consequently, if a practice is missing (\textit{\gls{llm} output = 0}), the overall score drops, reflecting the security principle that a system is only as strong as its weakest link~\cite{lemesTrustworthinessAssessmentWeb2019}.

The weights for this aggregation reflect the relative importance of each practice, derived empirically from real-world vulnerability data~\cite{lemesTrustworthinessAssessmentWeb2019}. For each \gls{owasp} practice, associated \gls{cwe} categories~\cite{cwe} are identified, and CVE occurrence counts from public databases (NVD, CVE Details) are used to compute importance coefficients. Practices linked to more frequently observed vulnerabilities (\textit{e.g.}, Practice~\#15 associated with SQL Injection) are assigned higher weights.

Given a function $f$ with practice assessments $\{r_1, \ldots, r_n\}$ where $r_i \in \{0,1\}$, the trustworthiness score is computed via the Generalized Conjunction/Disjunction (GCD) function with power mean $r=-1$ (``weak simultaneity'' operator \cite{lemesTrustworthinessAssessmentWeb2019}):

\begin{equation}
    \text{Trustworthiness}(f) = \left( \sum_{i=1}^{n} w_i \cdot r_i^{-1} \right)^{-1}
    \label{eq:trust_score}
\end{equation}

where $w_i$ are CVE-derived weights normalized to $\sum_{i=1}^{n} w_i = 1$. 

We compute reference trustworthiness scores for all 42 functions using ground truth labels and Equation~\ref{eq:trust_score}.
These scores serve as benchmarks for evaluating \gls{llm}-generated trustworthiness estimates and are intended for relative comparison and risk-based screening rather than absolute compliance assessment. This integration of the \gls{qm} demonstrates a complete pipeline from automated secure coding practice detection to quantitative trustworthiness assessment, addressing the scalability limitation identified by~\cite{lemesTrustworthinessAssessmentWeb2019}.

\section{Experimental Results and Discussion}
\label{sec:results_discussion}
This section presents and discusses the experimental results to answer the three RQs. We first analyze the baseline detection capabilities of \glspl{llm} under zero-shot conditions (\gls{rq}1), followed by the impact of different prompt enrichment strategies (\gls{rq}2). Finally, we assess whether \gls{llm}-based classifications can be reliably integrated into an automated Trustworthiness Score computation (\gls{rq}3). All reported metrics are macro-averaged across the three target classes (Followed, Not Followed, and NA), assigning equal weight to each class, ensuring that the aggregate score is not biased by the prevalence of the majority class in the dataset. 

Given the extensive nature of the experimental data, we provide a replication package on GitHub (\url{https://github.com/icstsubmission-prog/icst_submission_2026}) containing the full dataset, ground truth, prompts, and source code. Additionally, we offer an interactive platform (\url{https://icstsubmission-prog.github.io/icst_submission_2026/tool/src/dashboard_llms/dashboard/index.html}) for granular result visualizations, allowing for deeper exploration beyond the scope of this discussion.

\subsection{Prompt 1 (Baseline)}
\label{sec:res:baseline}
Table~\ref{tab:prompt1_metrics} reports the performance of each model in the identification of OWASP Input Validation secure coding practices~\cite{owaspSecureCodingPractices2025} using a zero-shot prompt without external context.

\begin{table}[htbp]
    \centering
    \caption{Overall Performance Metrics by Model (Prompt 1)}
    \label{tab:prompt1_metrics}
    \small
    \begin{tabular}{lcccc}
        \textbf{Model} & \textbf{M. Acc.} & \textbf{M. Prec.} & \textbf{M. Rec.} & \textbf{M. F1.} \\ 
        \midrule
        gpt-4.1 & 0.9494 & 0.9790 & 0.8509 & \textbf{0.8932} \\
        gpt-4.1-mini & 0.8824 & 0.8493 & 0.8290 & 0.8080 \\
        gemini-2.5-flash & 0.8452 & 0.8146 & 0.8001 & 0.7732 \\
        gpt-4o-mini & 0.6548 & 0.7380 & 0.6961 & 0.6092 \\ 
        gpt-3.5-turbo & 0.5476 & 0.5413 & 0.7121 & 0.5213 \\
        \bottomrule
    \end{tabular}
\end{table}

The results establish a clear performance hierarchy. The \textit{gpt-4.1} dominates with a Macro F1-score of 0.89, driven by Macro Precision $\approx$0.98, indicating that when the model flags a practice as secure or insecure, it is almost certainly correct, a critical trait for automated auditing. In the mid-tier, \textit{gpt-4.1-mini} and \textit{gemini-2.5-flash} demonstrate that lightweight models can effectively identify common patterns, achieving F1-scores of $\approx$0.81 and $\approx$0.77, respectively. However, a significant performance gap separates these from \textit{gpt-3.5-turbo} (F1: $\approx$0.52), which exhibits high recall but poor precision ($\approx$0.54), suggesting classification based on superficial keywords rather than semantic understanding.

\begin{figure}[htbp]
    \centering
    \includegraphics[width=1\columnwidth]{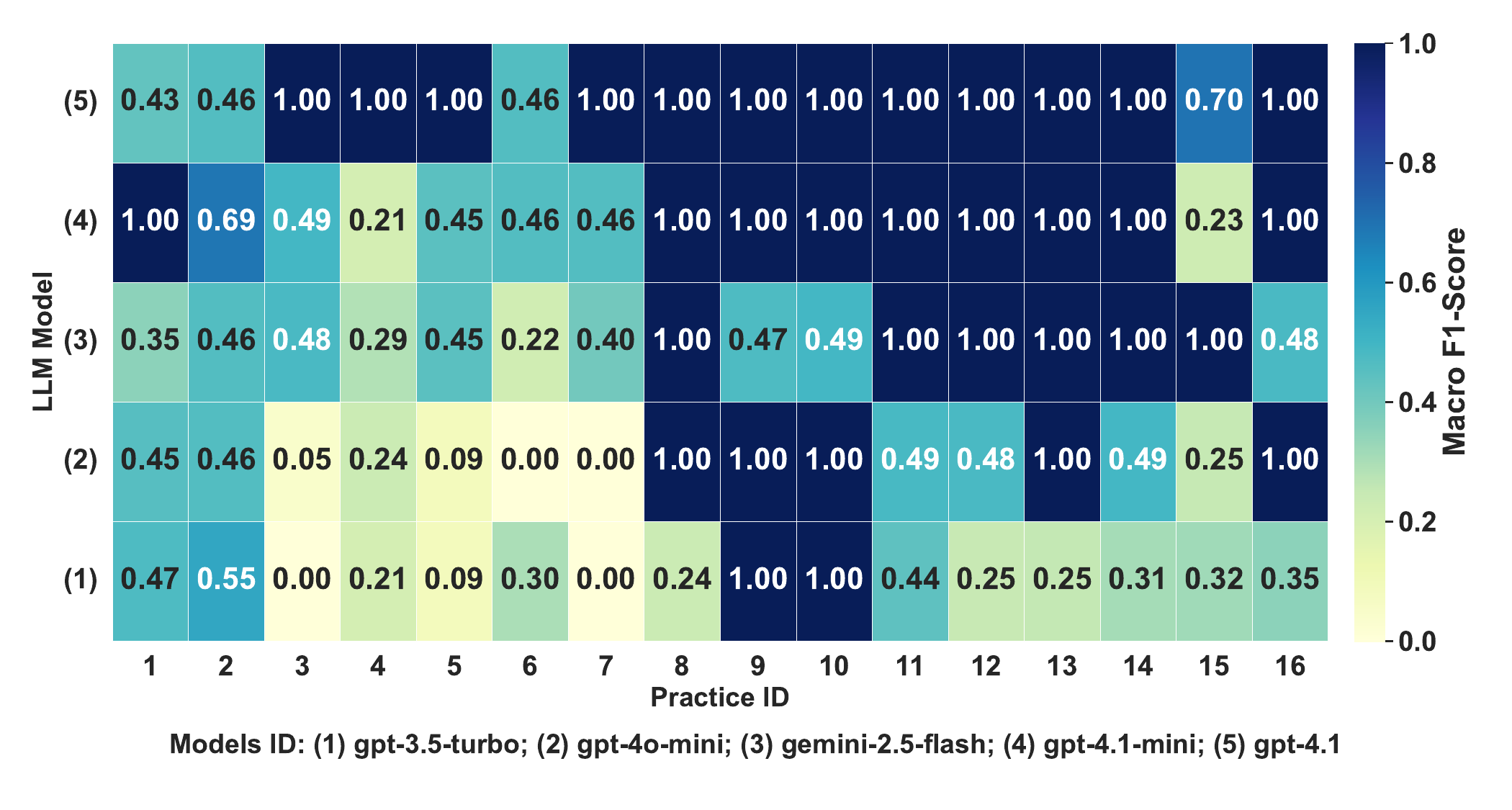}
    \caption{Macro F1-Score per Practice (Prompt 1).}
    \label{fig:heatmap_prompt1}
\end{figure}

Figure~\ref{fig:heatmap_prompt1} reveals that model performance is highly correlated with the nature of the security verification required. We identify four distinct failure modes:
\begin{itemize}
    \item \textbf{Practice~6 (Reject Failed Validation):} This practice requires verifying whether execution terminates after a validation failure (\textit{e.g.}, through a \texttt{return} or \texttt{throw}). Several models detected the presence of a validation condition but failed to verify whether execution was halted, resulting in misclassification when vulnerable code paths remained reachable. This outcome reflects limitations in analyzing control-flow semantics.

    \item \textbf{Practices~2 (Classify/Validate Sources),~3 (Centralized Routine), and~5 (Canonicalize Enconding):} These practices depend on the semantic role of variables. When the prompt does not specify whether a variable corresponds to raw user input or to a trusted identifier, models failed to determine the adequacy of validation logic. In these instances, \textit{gpt-3.5-turbo} frequently classified insufficient validation as secure, whereas \textit{gpt-4.1} tended to classify ambiguous cases as insecure.

    \item \textbf{Practices~4 (Specify Character Sets) \&~7 (Validate After Decode):} These practices require verifying character encoding enforcement. In Java applications, encoding configuration is typically handled outside the local function scope (\textit{e.g.}, via framework-level configuration). Since Prompt~1 limits the analysis to isolated code snippets, models cannot observe such external mechanisms, leading to inconsistent classification.

    \item \textbf{Practice~15 (Hazardous Char Controls):} This practice requires tracking conditional dependencies between hazardous input acceptance and mitigation mechanisms. \textit{gemini-2.5-flash} achieved an F1-score of 1.0, while several GPT-based models obtained lower scores. This indicates variability in the ability to resolve conditional cause–effect relations under zero-shot conditions.
\end{itemize}

\textbf{\gls{rq}1: Can large language models identify adherence to OWASP Input Validation secure coding practices under baseline (zero-shot) prompting conditions?}
Results for \gls{rq}1 show that baseline \glspl{llm} can partially identify \gls{owasp} practices using only function-level code, though effectiveness varies by requirement type. Performance is highest for practices with explicit syntactic patterns or local checks. Conversely, practices requiring control-flow reasoning, data provenance, or architectural context exhibit significantly lower accuracy. This demonstrates that baseline \gls{llm}-based analysis is feasible for locally verifiable practices but is insufficient for comprehensive security assessment in the absence of additional contextual guidance.

\subsection{Prompt 2 (CWE + Bad Examples)}
\label{sec:res:prompt2}
Table~\ref{tab:prompt2_metrics} reports the performance of each model when the baseline prompt is augmented with \gls{cwe} mappings and negative code examples.

\begin{table}[htbp]
    \centering
    \caption{Overall Performance Metrics by Model (Prompt 2 vs. Baseline)}
    \label{tab:prompt2_metrics}
    \small
    \begin{tabular}{lcccc}
        \textbf{Model} & \textbf{M. Acc.} & \textbf{M. Prec.} & \textbf{M. Rec.} & \textbf{M. F1.} \\
        \midrule
        gpt-4.1 & \up{0.961} & \down{0.960} & \up{0.905} & \textbf{\up{0.930}} \\
        gpt-4.1-mini & \down{0.850} & \down{0.764} & \up{0.852} & \down{0.790} \\
        gemini-2.5-flash & \down{0.802} & \down{0.657} & \down{0.731} & \down{0.679} \\
        gpt-4o-mini & \up{0.751} & \down{0.722} & \up{0.756} & \up{0.691} \\
        gpt-3.5-turbo & \up{0.573} & \down{0.464} & \down{0.580} & \down{0.474} \\
        \bottomrule
    \end{tabular}
    \footnotesize{\\ \vspace{2pt} Note: \textcolor{green!60!black}{$\uparrow$} and \textcolor{red}{$\downarrow$} denote increase or decrease relative to Baseline (Prompt 1).}
\end{table}

The introduction of definitions \gls{cwe} and negative examples produces divergent effects (Table~\ref{tab:prompt2_metrics}). \textit{gpt-4.1} takes advantage of this context to achieve a score of Macro F1 (0.93), with improvements in both accuracy (+0.0116) and recall (+0.054). However, this complexity imposes costs on other architectures. The \textit{gemini-2.5-flash} shows regression compared to the baseline (F1 drops from $\approx$0.77 to $\approx$0.68), with difficulty integrating dense prompt information. Similarly, \textit{gpt-3.5-turbo} suffers from contextual overload, with F1 falling to $\approx$0.47, confirming that simpler models fail to extract relevant instructions when saturated with informational noise.

\begin{figure}[htbp]
    \centering
    \includegraphics[width=1\columnwidth]{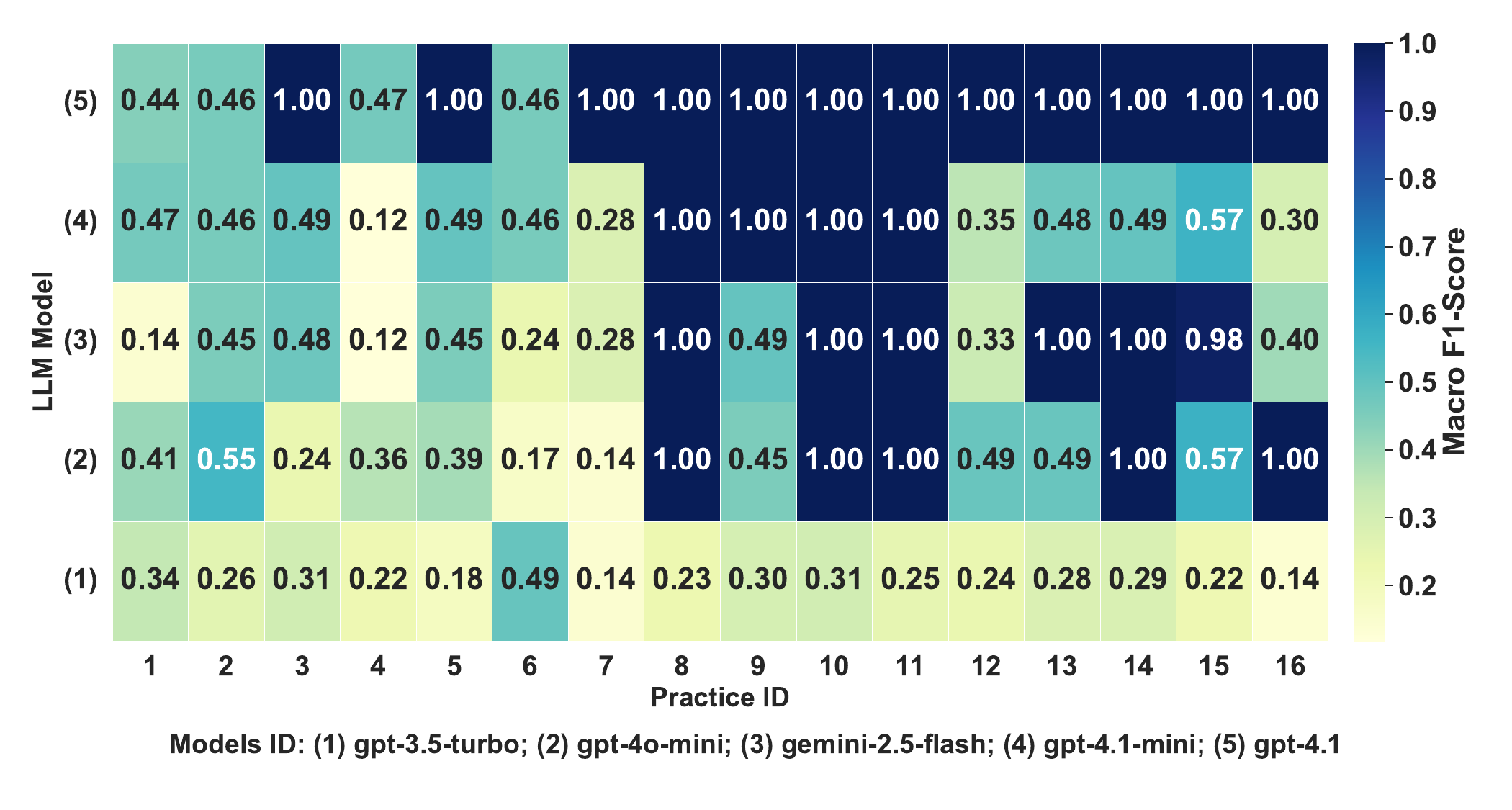}
    \caption{Heatmap of Macro F1-Scores per Practice (Prompt 2).}
    \label{fig:heatmap_prompt2}
\end{figure}

Figure~\ref{fig:heatmap_prompt2} illustrates practice-level performance for Prompt~2, highlighting three distinct behaviors:

\begin{itemize}
    \item \textbf{Practices~3 \&~5:} Negative examples improved performance across all models by providing explicit patterns of violation. This enabled the detection of non-compliant structures previously classified as safe or not applicable.

    \item \textbf{Practices~4 \&~7, 9 (ASCII Header), and~12 (Data Range):} Performance declines for practices dependent on external configuration or implicit context. The high specificity of the negative examples led to overfitting, where models failed to identify violations that did not strictly match the provided patterns. This effect was most pronounced in intermediate models, indicating limited generalization capability beyond the illustrated examples.

    \item \textbf{Practice~15:} Models with higher reasoning capacity utilized negative examples to associate hazardous inputs with missing mitigation logic. Lower-capacity models showed no consistent improvement, indicating that the added prompt complexity did not translate into more accurate classification for logic-dependent tasks.
\end{itemize}

These results indicate that few-shot prompting improves pattern recognition for local violations but fails to address architectural context or semantic ambiguity.

\subsection{Prompt 3 (Function Call Context)}
\label{sec:results:prompt3}
Prompt 3 augmented the analysis by retrieving and appending the source code of referenced helper functions (via Joern \cite{joernDocs}) to provide structural context. \textit{Note: The \gls{llm} gemini-2.5-flash is excluded from this analysis as the expanded prompt size consistently exceeded the model's token limit.}

Table~\ref{tab:prompt3_metrics} details the performance metrics. The data indicates a general regression in Macro F1-Scores compared to the baseline, primarily attributed to a reduction in recall capabilities.

\begin{table}[htbp]
    \centering
    \caption{Overall Performance Metrics by Model (Prompt 3 vs. Baseline)}
    \label{tab:prompt3_metrics}
    \small
    \begin{tabular}{lcccc}
        \textbf{Model} & \textbf{M. Acc} & \textbf{M. Prec.} & \textbf{M. Rec.} & \textbf{M. F1.} \\ 
        \midrule
        gpt-4.1 & \down{0.9048} & \textbf{\down{0.9565}} & \down{0.7272} & \textbf{\down{0.7474}} \\
        gpt-4.1-mini & \down{0.8050} & \down{0.7516} & \textbf{\down{0.7767}} & \down{0.7192} \\
        gpt-4o-mini & \down{0.6495} & \down{0.3921} & \down{0.5629} & \down{0.4187} \\
        gpt-3.5-turbo & \down{0.3824} & \down{0.3794} & \down{0.4383} & \down{0.3172} \\
        \bottomrule
    \end{tabular}
    \footnotesize{\\ \vspace{2pt} Note: \textcolor{green!60!black}{$\uparrow$} and \textcolor{red}{$\downarrow$} denote increase or decrease relative to Baseline (Prompt 1).}
\end{table}

Contrary to the hypothesis that providing auxiliary function code would enhance verification, Prompt 3 resulted in an overall performance regression (Table~\ref{tab:prompt3_metrics}). The \textit{gpt-4.1} declined from an F1 of $\approx$0.89 (Prompt 1) to $\approx$0.75, a shift driven by a significant decrease in recall ($\approx$0.73 vs $\approx$0.85). Although precision remained robust (0.96), the inclusion of auxiliary functions increased informational noise, hindering the effective identification of valid safety implementations (false negatives). 

\begin{figure}[htbp]
    \centering
    \includegraphics[width=1\columnwidth]{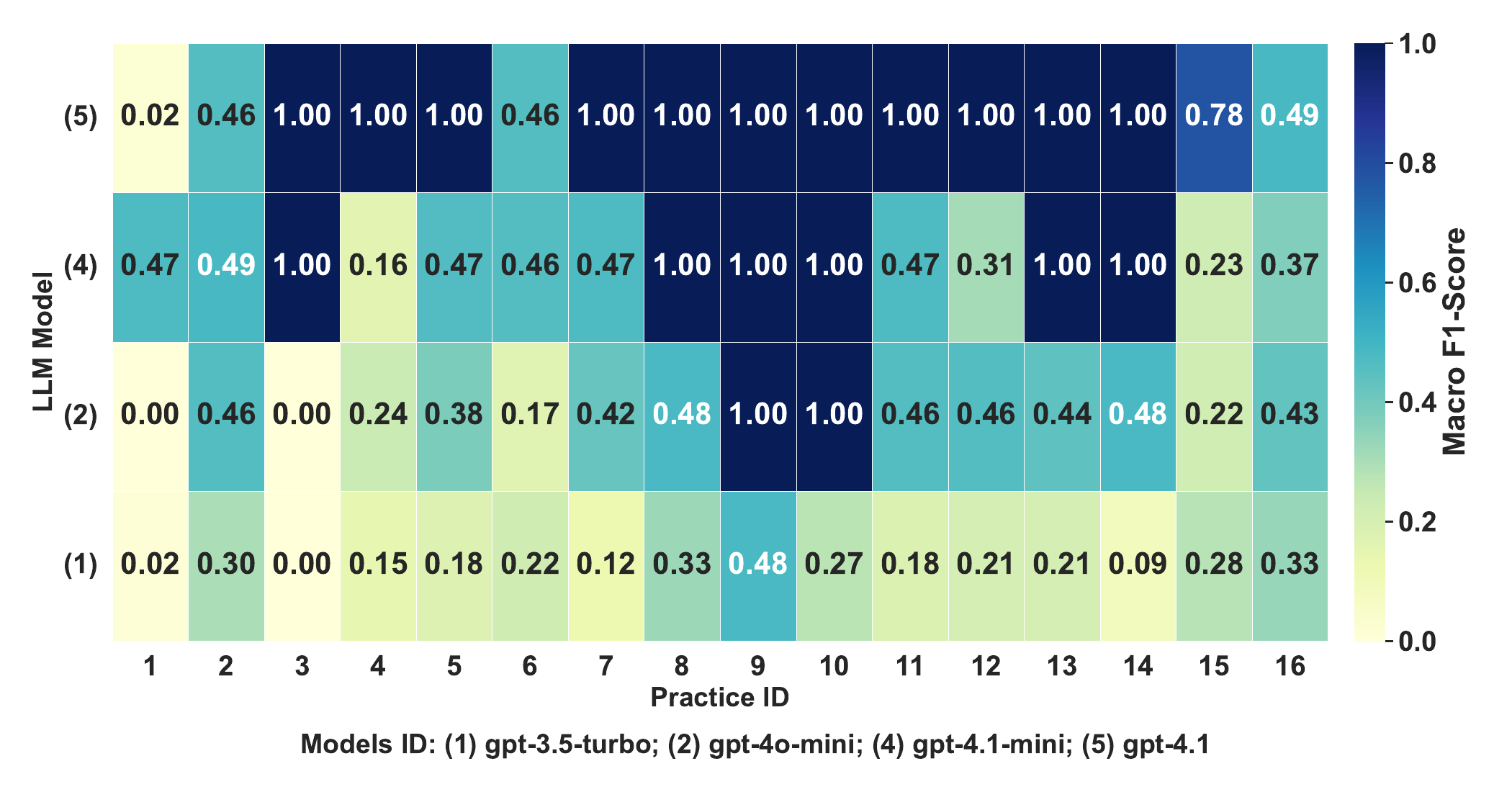}
    \caption{Heatmap of Macro F1-Scores per Practice (Prompt 3).}
    \label{fig:heatmap_prompt3}
\end{figure}

Figure~\ref{fig:heatmap_prompt3} illustrates this performance inconsistency. Unlike the uniform high-performance clusters observed in Prompt~2, Prompt~3 exhibits dispersed results:
\begin{itemize}
    \item \textbf{High-capacity models:} Even \textit{gpt-4.1} (Row 5) displays irregular performance dips in foundational practices such as Practice~1 (Trusted System). The heatmap suggests that the increased code volume impeded the model's ability to correctly correlate security requirements with their specific implementations within the auxiliary functions, resulting in missed detections.

    \item \textbf{Mid-tier models:} The \textit{gpt-4.1-mini} (Row 4), despite its general capability, shows significant regression in specific domains, such as Practice~4 (Charset), where detection rates decreased. This supports the observation that without a mechanism to filter relevant code segments, the expanded context acts as noise, obscuring vulnerabilities rather than clarifying them.

    \item \textbf{Lower-tier models:} Models with lower reasoning capacity (Rows 1 and 2) display consistently low scores. \textit{gpt-3.5-turbo} and \textit{gpt-4o-mini} fail to register valid detections for the majority of practices, suggesting an architectural bottleneck when processing high-density prompts.
\end{itemize}

These results indicate that naively extending the analysis context by injecting raw dependencies is insufficient and may negatively affect classification reliability.

\subsection{Prompt 4 (Rule-Based)}
\label{sec:results:prompt4}

Table~\ref{tab:prompt4_metrics} presents the results for Prompt~4, which enriches the baseline structure with explicit, step-by-step instructions.

\begin{table}[htbp]
    \centering
    \caption{Overall Performance Metrics by Model (Prompt 4 vs. Baseline)}
    \label{tab:prompt4_metrics}
    \small
    \begin{tabular}{lcccc}
        \textbf{Model} & \textbf{M. Acc} & \textbf{M. Prec.} & \textbf{M. Rec.} & \textbf{M. F1.} \\ 
        \midrule
        gpt-4.1 & \down{0.9360} & \down{0.8889} & \up{0.9272} & \textbf{\up{0.9020}} \\
        gpt-4.1-mini & \up{0.8854} & \down{0.8461} & \up{0.8789} & \up{0.8397} \\
        gemini-2.5-flash & \up{0.8720} & \down{0.7818} & \down{0.7946} & \up{0.7744} \\
        gpt-4o-mini & \up{0.8438} & \up{0.8191} & \up{0.7785} & \up{0.7380} \\ 
        gpt-3.5-turbo & \down{0.5327} & \up{0.7068} & \down{0.6536} & \up{0.5451} \\
        \bottomrule
    \end{tabular}
    \footnotesize{\\ \vspace{2pt} Note: \textcolor{green!60!black}{$\uparrow$} and \textcolor{red}{$\downarrow$} denote increase or decrease relative to Baseline (Prompt 1).}
\end{table}

The rule-based request produces robust performance (Table~\ref{tab:prompt4_metrics}), with \textit{gpt-4.1} achieving a Macro F1 of $\approx$0.90. Although this represents a slight decrease for the top-level model compared to Prompt 2 (0.93), structured constraints have shown to be highly beneficial for smaller architectures. Notably, the efficient GPT models surpassed their few-shot performance: \textit{gpt-4.1-mini} improved to $\approx$0.84 (vs. 0.79 in Prompt 2) and \textit{gpt-4o-mini} rose to $\approx$0.74 (vs. 0.69 in Prompt 2). This indicates that explicit \enquote{if-then} directives effectively compensate for the lower reasoning capacity of compact models, offering a more token-efficient alternative to few-shot examples.

\begin{figure}[htbp]
    \centering
    \includegraphics[width=1\columnwidth]{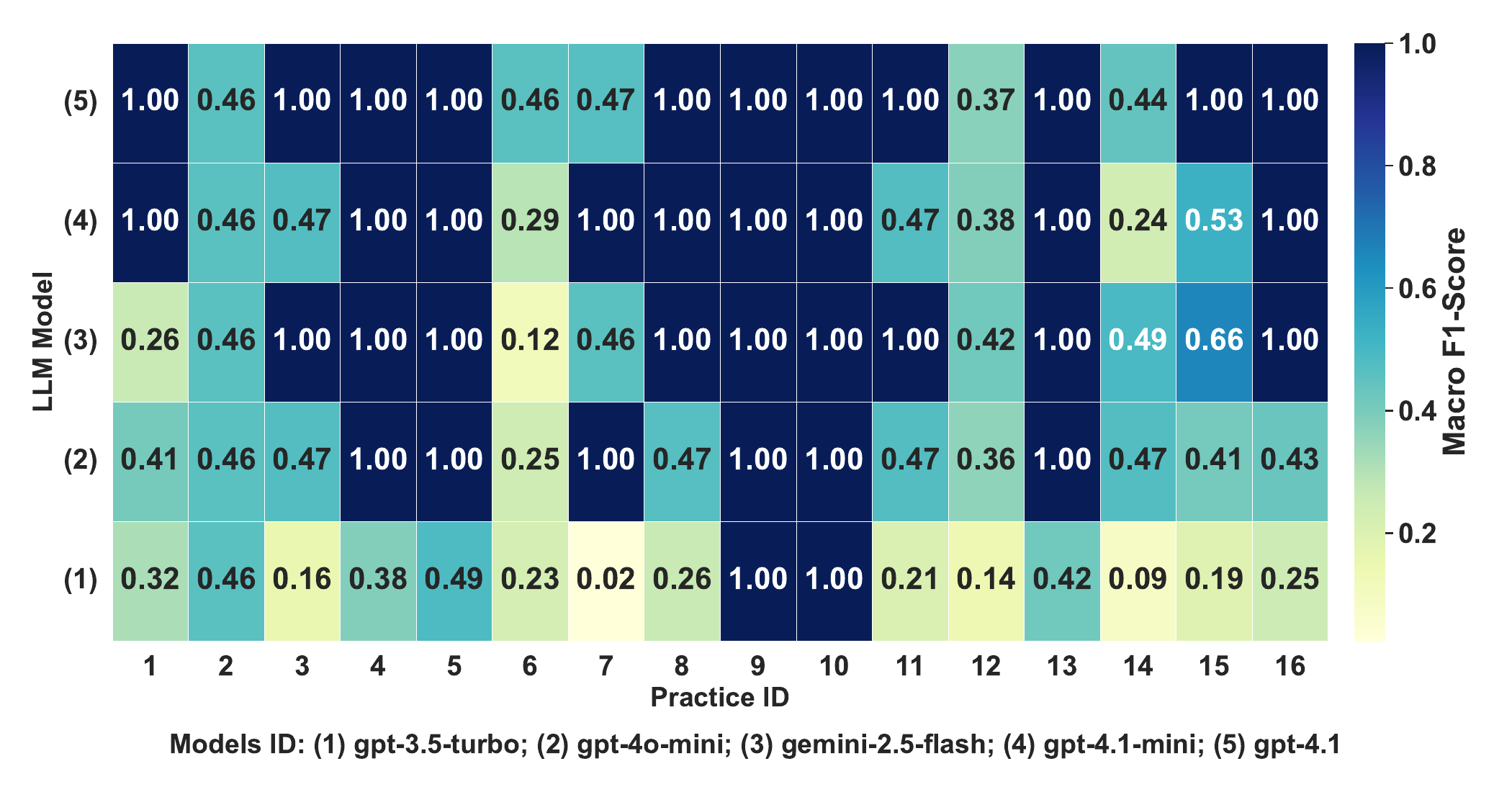}
    \caption{Heatmap of Macro F1-Scores per Practice (Prompt 4).}
    \label{fig:heatmap_prompt4}
\end{figure}

As Figure~\ref{fig:heatmap_prompt4} shows, explicit rules produce varying results depending on the characteristics of the practice:
\begin{itemize}
    \item \textbf{Practice~15:} Explicit rules regarding conditional logic significantly improved the performance of GPT-family models compared to the baseline (\textit{e.g.}, \textit{gpt-4.1-mini} rise from 0.23 to 0.53). Although this score is slightly lower than the peak achieved with few-shot examples in Prompt~2 (0.57), it validates that textual rules can effectively guide models to verify mitigation branches, bridging the gap observed in the zero-shot baseline.

    \item \textbf{Practices~4 \&~7:} Performance improved because the explicit rules redefined the verification scope. By instructing models to validate local function parameters for charset definitions, instead of attempting to infer missing global configurations, the rules established a verifiable local proxy for compliance. This operational shift directly addressed the context blindness, raising the accuracy in Practice~4 to $\approx$92.38\% and notably boosting \textit{gpt-4.1-mini} F1-score in Practice~7 from $\approx$0.46 (Baseline) to 1.00.

    \item \textbf{Practices~12 \&~14:} Performance scores decreased relative to the baseline. For Data Range (P12), the rules lacked the domain-specific parameters required to validate numerical boundaries, resulting in misclassification. In Whitelisting (P14), the strict rule definitions led to false negatives, as models failed to recognize valid sanitization logic, such as complex expressions or utility methods, that deviated from the explicit patterns provided in the prompt.
\end{itemize}

Rule-based prompting improves consistency for locally verifiable practices but remains limited when validation depends on information outside the analyzed snippet.


\textbf{\gls{rq}2: What is the impact of different prompt context enrichment strategies on the accuracy and consistency of \gls{llm}-based secure coding practice identification?}
Regarding \gls{rq}2, the results show that prompt effectiveness is highly dependent on model architecture (Table~\ref{tab:best_prompt_gain}). While \textit{gpt-4.1} reached peak performance using semantic context (Prompt~2, F1 = $\Delta$+0.0368), smaller models \textit{gpt-4o-mini}, \textit{gemini-2.5-flash}, and \textit{gpt-3.5-turbo} required rule-based constraints (Prompt~4) for optimal results, with \textit{gpt-4o-mini} achieving the study's highest gain ($\Delta$+0.1288). Notably, Prompt~3 caused performance regression across all models. These findings indicate that Prompt~4 allowing lower-capacity models to approximate high-capacity accuracy.


\begin{table}[htbp]
    \centering
    \caption{Maximum Improvement by Model: Prompt 1 vs. Best Strategy (Macro F1)}
    \label{tab:best_prompt_gain}
    \small
    \begin{tabular}{lcccc}
        \textbf{Model} & \textbf{P.1 (F1)} & \textbf{Best P.} & \textbf{Best F1} & \textbf{Gain ($\Delta$)} \\
        \midrule
        \textit{gpt-4.1} & 0.8932 & P2 & 0.9300 & \textbf{+0.0368} \\
        \textit{gpt-4.1-mini} & 0.8080 & P4 & 0.8397 & \textbf{+0.0317} \\
        \textit{gemini-2.5-flash} & 0.7732 & P4 & 0.7744 & \textbf{+0.0009} \\
        \textit{gpt-4o-mini} & 0.6092 & P4 & 0.7380 & \textbf{+0.1288} \\
        \textit{gpt-3.5-turbo} & 0.5213 & P4 & 0.5451 & \textbf{+0.0238} \\
        \bottomrule
    \end{tabular}
\end{table}

\subsection{Trustworthiness Score Assessment}
\label{sec:results:quality_model}
To validate the integration of \glspl{llm} into the \gls{qm}, we compared two distinct paradigms: (1) \textbf{\gls{llm}-driven Quality Model} analytical aggregation, where binary classifications are deterministically processed via the \gls{lsp} aggregation formula; and (2) \textbf{Trustworthiness Estimation} (Prompt 5), where the model is tasked with implicitly synthesizing a scalar trustworthiness score.
As established in Section~\ref{sec:methodology}, the validity of these scores is benchmarked against a ground truth. This baseline is calculated by applying the same \gls{lsp} aggregation logic to the manually verified labels of the dataset, representing the ideal trustworthiness assessment for each function.

\textbf{LLM-driven Quality Model:}
\label{sec:res:global_risk}
We compute trustworthiness scores by analytically aggregating \gls{llm}-derived binary classifications using the \gls{lsp} methodology, enabling structured and interpretable system-level trust assessment.

\begin{figure}[htbp]
    \centering
    \includegraphics[width=1\columnwidth]{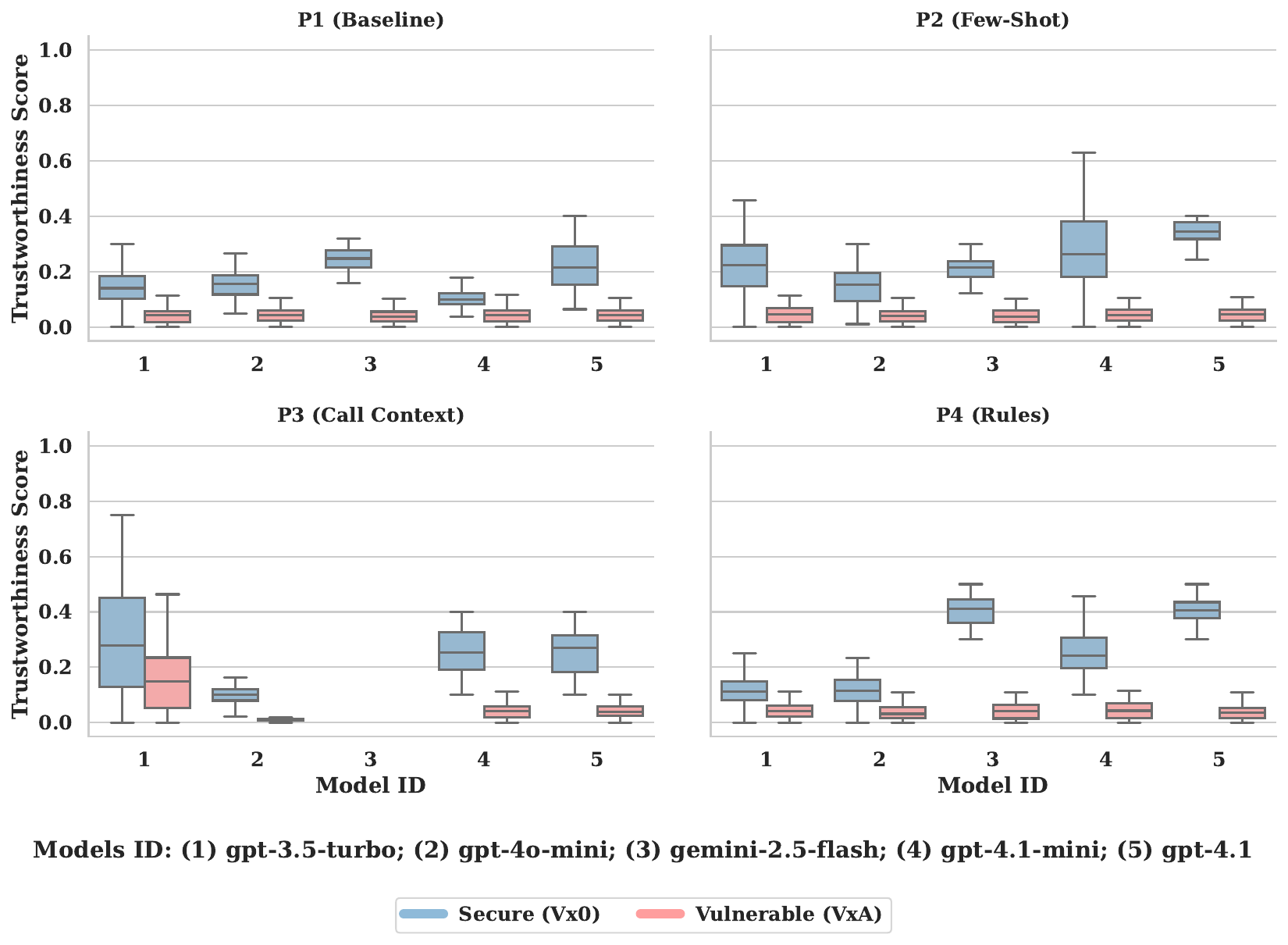}
    \caption{Evolution of Trustworthiness Scores across Prompt Strategies. Models ID: (1) gpt-3.5-turbo; (2) gpt-4o-mini; (3) gemini-2.5-flash; (4) gpt-4.1-mini; (5) gpt-4.1.}
    \label{fig:global_boxplot}
\end{figure}

To demonstrate the impact of prompt engineering on the final trustworthiness scores, Figure~\ref{fig:global_boxplot} presents the comparative distribution of scores across all 42 analyzed functions for the four experimental strategies. In these boxplots, the blue boxes represent secure code versions (expected high scores), while the red boxes represent vulnerable versions (expected low scores). 
A visual inspection reveals that scores are conservative (median $<0.40$). This is an expected consequence of the ``weak simultaneity'' operator ($r=-1$) in our \gls{qm}, which heavily penalizes the score if practice is missing. However, the critical metric is discriminatory capability.

The analysis highlights that adding raw context (Prompt~3) introduces significant aleatoric uncertainty, evident in the expanded Interquartile Ranges (IQR), particularly rendering \textit{gpt-3.5-turbo} unreliable. In contrast, the Rule-Based strategy (Prompt~4) proves to be the most robust. It minimizes dispersion and establishes a clear statistical separation between secure and vulnerable distributions, with no interquartile overlap. Crucially, Prompt~4 enables lightweight models like \textit{gpt-4.1-mini} and \textit{gemini-2.5-flash} to approximate the discriminative performance of high-capacity architectures (\textit{gpt-4.1} in Prompt~2), effectively compensating for the reasoning instability observed in Prompt~3.

To formally validate the models' ability to differentiate between secure and vulnerable functions, we performed a two-sided Mann-Whitney U test. The results confirm statistical significance ($p < 0.05$) for most configurations, indicating discriminatory ability. Conversely, smaller models in Prompt~3 show high $p$ values (\textit{e.g.}, $p \geq 0.71$ for \textit{gpt-3.5-turbo} and \textit{gpt-4o-mini}), indicating no statistically significant differences, confirming the visual evidence of high uncertainty and the loss of discriminatory power when relying on unstructured context.

\begin{figure}[htbp]
    \centering
    \includegraphics[width=1\columnwidth]{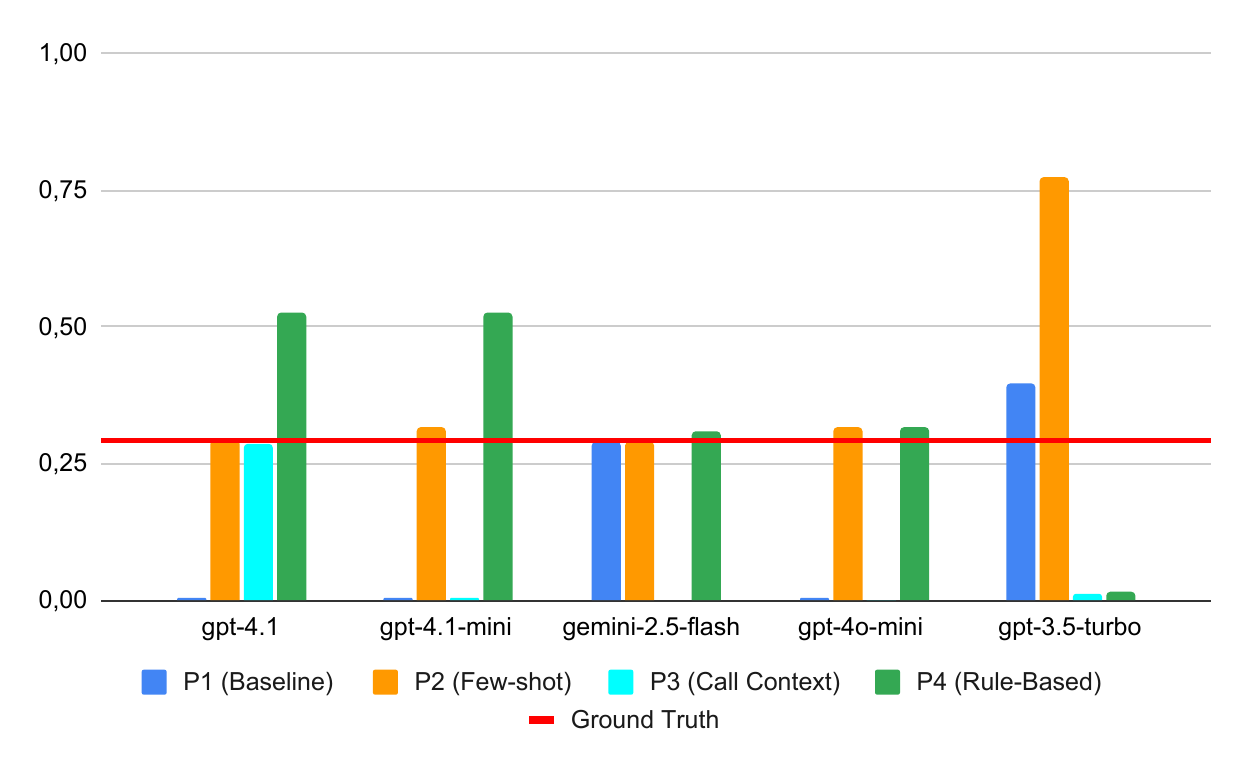}
    \vspace{-8pt}
    \caption{Function \texttt{enterAddress} - Secure Version (Vx0).}
    \label{fig:vxo_analysis}
\end{figure}

\begin{figure}[htbp]
    \centering
    \includegraphics[width=1\columnwidth]{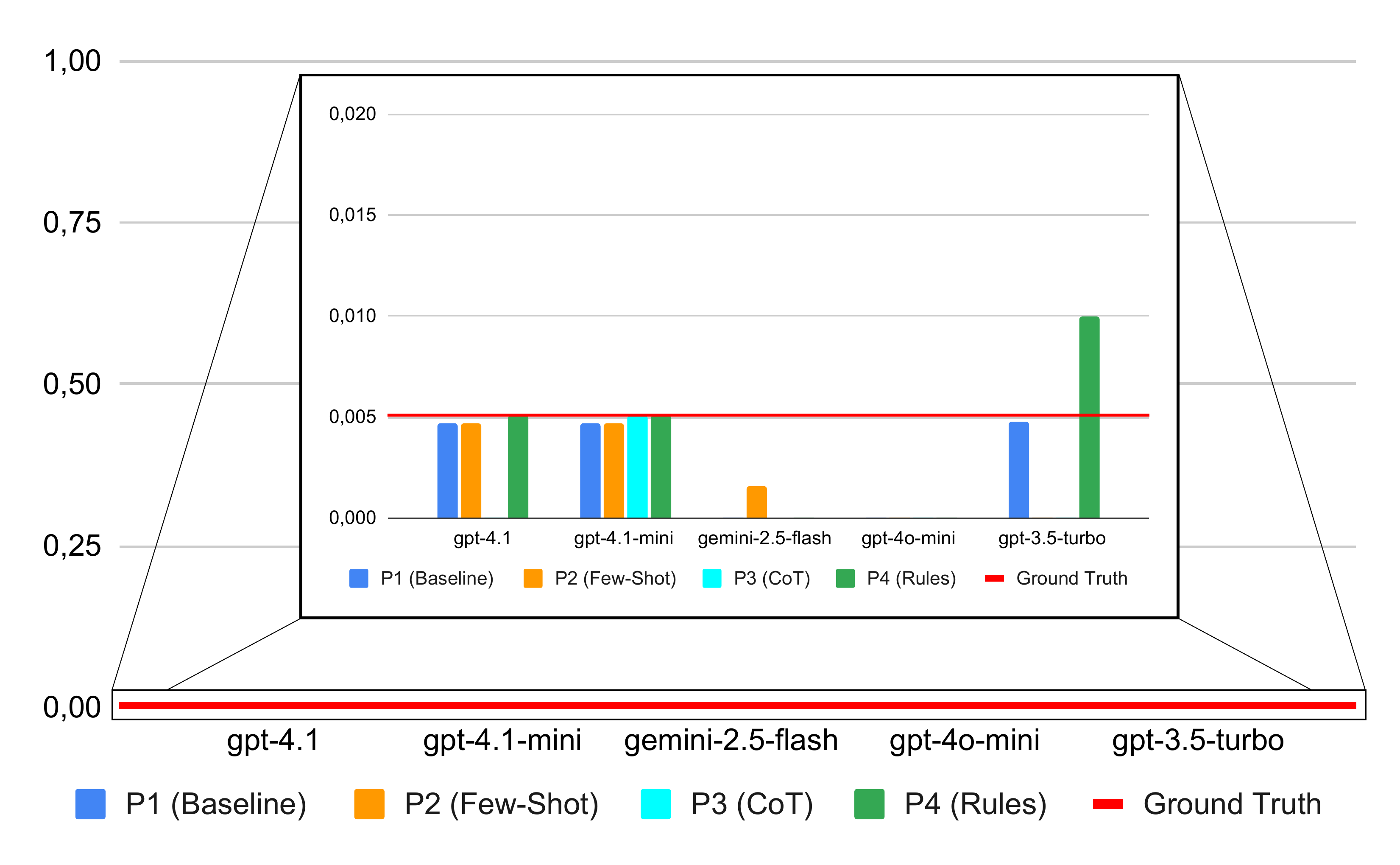}
    \vspace{-8pt}
    \caption{Function \texttt{enterAddress} - Vulnerable Version (VxA) with Zoom.}
    \label{fig:vxa_analysis}
\end{figure}

More important than the absolute value of the score is the differential capacity: can the model assign high scores to secure code and low scores to vulnerable code? Differences between scores should be interpreted in relative terms, enabling prioritization and comparison across implementations. To provide a concrete illustrative example of the sensitivity of the aggregated scores to code variations, we compare the secure and vulnerable versions of the \texttt{enterAddress} function as a representative case study. Figures~\ref{fig:vxo_analysis} and~\ref{fig:vxa_analysis} show the resulting trustworthiness scores for the secure version (\texttt{Vx0}) and the vulnerable version (\texttt{VxA}), respectively.

Visual comparison shows that \textit{gpt-3.5-turbo} produces artificially inflated scores in Prompt~1 and Prompt~2, indicating false positives that propagate through aggregation. Although Prompt~4 eliminates these hallucinations, it also suppresses the model's detection capability. In contrast, \textit{gemini-2.5-flash} demonstrates high stability, maintaining consistent alignment with the reference in Prompt~1, Prompt~2, and Prompt~4.

The \textit{gpt-4} family shows a critical dependence on structure. Although largely inactive in Prompt~1 and Prompt~3, the Prompt~4 produces scores that are not aligned with the ground truth. This confirms that while structured prompting improves reliability, high-capacity models can become slightly more aggressive in classification when rules are strictly enforced.

It is important to note that this improvement occurs without compromising the evaluation of insecure implementations; in the vulnerable version (\texttt{VxA}), all models consistently maintain scores close to zero (0.00-0.01) on effective prompts.

\textbf{Trustworthiness Estimation:}
\label{sec:res:direct_score}
We also evaluated the ability of \glspl{llm} to directly generate trustworthiness scores in the interval $[0,1]$, without relying on an explicit aggregation model. Across all evaluated models, the results exhibited substantial deviations from the reference scores, with \gls{mae} values ranging from 0.40 to 0.60. The generated scores showed high variability across equivalent inputs and did not consistently reflect the relative importance of individual secure coding practices. These results indicate that, while direct score generation is feasible, it is not sufficiently reliable to serve as a standalone mechanism for trustworthiness assessment in this setting.

\textbf{\gls{rq}3: To what extent can \glspl{llm} be used to produce reliable trustworthiness scores?}
Results for \gls{rq}3 reveal that reliability depends strictly on the architectural approach. \gls{llm}-Driven estimation proved inadequate, as \glspl{llm} failed to implicitly synthesize multi-criteria assessments or reflect practice importance. Conversely, analytical aggregation via the \gls{lsp} \gls{qm} delivered stable, discriminative metrics, ensuring the final score is a deterministic function of detected practices rather than a stochastic generation. By automating practice detection, \glspl{llm} overcome the scalability limitations of manual, expertise-intensive assessments. We conclude that while \glspl{llm} are insufficient for direct quantification, they enable reliable, scalable trustworthiness assessment when serving as automated inputs to a structured and explicit \gls{qm}.

\section{Threats to Validity}
\label{sec:threats}

Like any empirical study, this work is subject to limitations that must be considered when interpreting the results.

\textbf{Construct Validity:} The main threat lies in the definition of the ground truth. The classification of security practices relied on manual analysis subject to human interpretation of secure coding standards \cite{owaspSecureCodingPractices2025}.

\textbf{Internal Validity:} This is affected by the restriction to proprietary model architectures (GPT and Gemini families). The exclusion of open-source models limits our understanding of whether phenomena such as the reasoning collapse in prompts are universal. Additionally, the stochastic and \textit{black box} nature of \glspl{llm} prevents a complete audit of the decision-making process, meaning we cannot guarantee that correct answers result from valid safety reasoning rather than spurious statistical correlations.

\textbf{External Validity:} The generalization of results is limited by the scope of the dataset, which focuses exclusively on Java and input validation for SQL injection prevention within the WSVD-Bench application \cite{wsvd-bench}. Consequently, conclusions may not directly extrapolate to dynamic languages or other secure coding practices \cite{owaspSecureCodingPractices2025}.


\section{Conclusion and Future Work}
\label{sec:conclusion}
Trustworthiness is a critical requirement for modern web applications operating in security- and safety-sensitive contexts, yet its assessment remains challenging due to the reliance of existing approaches on manual, expertise-intensive verification. This study demonstrates that \glspl{llm} can effectively automate secure coding practice assessment and enable scalable trustworthiness evaluation when combined with a structured and deterministic \gls{qm}. This is especially relevant in modern \gls{ci/cd} environments, where trustworthiness must be assessed continuously under tight development timelines.


Future work will extend the proposed approach to additional programming languages and vulnerabilities. Another promising direction is the selective integration of retrieval-based mechanisms to overcome context limits, as well as embedding automated trustworthiness assessment into \gls{ci/cd} pipelines for continuous evaluation in modern development workflows.

\section*{Acknowledgment}
Content produced within the scope of the Agenda "NEXUS - Pacto de Inovação - Transição Verde e Digital para Transportes, Logística e Mobilidade", financed by the Portuguese Recovery and Resilience Plan (PRR), with no. C645112083-00000059 (investment project no. .º 53). It was also developed within the scope of the research undertaken by “Projeto n.º 2024.07660.IACDC”, and by FCT I.P. research unit UID/00326. This work used computing resources provided through the Google Cloud Research Credits program with the award GCP19980904.

\bibliographystyle{ieeetr}
\bibliography{new}

\begin{thebibliography}{10}

\bibitem{antunesDefendingWebApplication2012}
N.~Antunes and M.~Vieira, ``Defending against {{Web Application
  Vulnerabilities}},'' {\em Computer}, vol.~45, pp.~66--72, Feb. 2012.

\bibitem{zhuSurveySecurityAnalysis2024}
H.~Zhu, L.~Yang, L.~Wang, and V.~S. Sheng, ``A {{Survey}} on {{Security
  Analysis Methods}} of {{Smart Contracts}},'' {\em IEEE Transactions on
  Services Computing}, vol.~17, pp.~4522--4539, Nov. 2024.

\bibitem{faisalfadlallaInputValidationVulnerabilities2023}
F.~Faisal~Fadlalla and H.~T. Elshoush, ``Input {{Validation Vulnerabilities}}
  in {{Web Applications}}: {{Systematic Review}}, {{Classification}}, and
  {{Analysis}} of the {{Current State-of-the-Art}},'' {\em IEEE Access},
  vol.~11, pp.~40128--40161, 2023.

\bibitem{lakshmananNewCriticalMOVEit2023}
R.~Lakshmanan, ``New {{Critical MOVEit Transfer SQL Injection Vulnerabilities
  Discovered}} - {{Patch Now}}!.''
  \url{https://thehackernews.com/2023/06/new-critical-moveit-transfer-sql.html},
  June 2023.
\newblock Accessed: 2025-01-25.

\bibitem{owaspTop10}
OWASP, ``Owasp top 10: The ten most critical web application security risks.''
  \url{https://owasp.org/www-project-top-ten/}, 2023.
\newblock Accessed: 2025-01-11.

\bibitem{mitropoulosDefendingWebApplication2019}
D.~Mitropoulos, P.~Louridas, M.~Polychronakis, and A.~D. Keromytis, ``Defending
  {{Against Web Application Attacks}}: {{Approaches}}, {{Challenges}} and
  {{Implications}},'' {\em IEEE Transactions on Dependable and Secure
  Computing}, vol.~16, pp.~188--203, Mar. 2019.

\bibitem{ibrahimkhalafWebAttackDetection2021}
O.~Ibrahim~Khalaf, M.~Sokiyna, Y.~Alotaibi, A.~Alsufyani, and S.~Alghamdi,
  ``Web {{Attack Detection Using}} the {{Input Validation Method}}: {{DPDA
  Theory}},'' {\em Computers, Materials \& Continua}, vol.~68, no.~3,
  pp.~3167--3184, 2021.

\bibitem{sharPredictingCommonWeb2012}
L.~K. Shar and H.~B.~K. Tan, ``Predicting common web application
  vulnerabilities from input validation and sanitization code patterns,'' in
  {\em Proceedings of the 27th {{IEEE}}/{{ACM International Conference}} on
  {{Automated Software Engineering}}}, {{ASE}} '12, (New York, NY, USA),
  pp.~310--313, Association for Computing Machinery, Sept. 2012.

\bibitem{scholtePreventingInputValidation2012}
T.~Scholte, W.~Robertson, D.~Balzarotti, and E.~Kirda, ``Preventing {{Input
  Validation Vulnerabilities}} in {{Web Applications}} through {{Automated Type
  Analysis}},'' in {\em 2012 {{IEEE}} 36th {{Annual Computer Software}} and
  {{Applications Conference}}}, pp.~233--243, July 2012.

\bibitem{Pereira2023}
J.~D. Pereira and M.~Vieira, ``An approach to characterize the security of
  open-source functions using lsp,'' in {\em 2023 IEEE 34th International
  Symposium on Software Reliability Engineering (ISSRE)}, pp.~137--147, 2023.

\bibitem{DAbruzzoPereira2020}
J.~D'Abruzzo~Pereira and M.~Vieira, ``{On the Use of Open-Source C/C++ Static
  Analysis Tools in Large Projects},'' in {\em 2020 16th European Dependable
  Computing Conference (EDCC)}, pp.~97--102, 2020.

\bibitem{singhAnalysisWebApplication2024}
R.~Singh, M.~Kumar~Gupta, D.~R. Patil, and S.~Maruti~Patil, ``Analysis of {{Web
  Application Vulnerabilities}} using {{Dynamic Application Security
  Testing}},'' in {\em 2024 {{IEEE}} 9th {{International Conference}} for
  {{Convergence}} in {{Technology}} ({{I2CT}})}, pp.~1--6, Apr. 2024.

\bibitem{zhuComprehensiveStudyStatic2024}
J.~Zhu, K.~Li, S.~Chen, L.~Fan, J.~Wang, and X.~Xie, ``A {{Comprehensive
  Study}} on {{Static Application Security Testing}} ({{SAST}}) {{Tools}} for
  {{Android}},'' {\em IEEE Transactions on Software Engineering}, vol.~50,
  pp.~3385--3402, Dec. 2024.

\bibitem{owasp2024}
OWASP, ``Interactive application security testing (iast).''
  \url{https://owasp.org/www-project-devsecops-guideline/latest/02c-Interactive-Application-Security-Testing},
  2024.
\newblock Accessed: 2025-01-11.

\bibitem{snykiast2024}
Snyk, ``Interactive application security testing (iast).''
  \url{https://snyk.io/pt-BR/articles/application-security/iast-interactive-application-security-testing/},
  2024.
\newblock Accessed: 2025-01-11.

\bibitem{lemesTrustworthinessAssessmentWeb2019}
C.~I. Lemes, V.~Naessens, and M.~Vieira, ``Trustworthiness {{Assessment}} of
  {{Web Applications}}: {{Approach}} and {{Experimental Study}} using {{Input
  Validation Coding Practices}},'' in {\em 2019 {{IEEE}} 30th {{International
  Symposium}} on {{Software Reliability Engineering}} ({{ISSRE}})},
  pp.~435--445, Oct. 2019.

\bibitem{owaspSecureCodingPractices2025}
O.~Foundation, ``Owasp secure coding practices quick reference guide.''
  \url{https://owasp.org/www-project-secure-coding-practices-quick-reference-guide/stable-en/},
  2025.

\bibitem{Vieira2025}
M.~Vieira, ``{ Leveraging LLMs for Trustworthy Software Engineering: Insights
  and Challenges },'' {\em Computer}, vol.~58, pp.~79--90, July 2025.

\bibitem{xuSystematicEvaluationLarge2022}
F.~F. Xu, U.~Alon, G.~Neubig, and V.~J. Hellendoorn, ``A systematic evaluation
  of large language models of code,'' in {\em Proceedings of the 6th {{ACM
  SIGPLAN International Symposium}} on {{Machine Programming}}}, {{MAPS}} 2022,
  (New York, NY, USA), pp.~1--10, Association for Computing Machinery, June
  2022.

\bibitem{zhaoSurveyLargeLanguage2024}
W.~X. Zhao, K.~Zhou, J.~Li, T.~Tang, X.~Wang, Y.~Hou, Y.~Min, B.~Zhang,
  J.~Zhang, Z.~Dong, Y.~Du, C.~Yang, Y.~Chen, Z.~Chen, J.~Jiang, R.~Ren, Y.~Li,
  X.~Tang, Z.~Liu, P.~Liu, J.-Y. Nie, and J.-R. Wen, ``A {{Survey}} of {{Large
  Language Models}},'' Oct. 2024.

\bibitem{bezziLargeLanguageModels2024}
M.~Bezzi, ``Large {{Language Models}} and {{Security}},'' {\em IEEE Security \&
  Privacy}, vol.~22, pp.~60--68, Mar. 2024.

\bibitem{fuChatGPTVulnerabilityDetection2023}
M.~Fu, C.~K. Tantithamthavorn, V.~Nguyen, and T.~Le, ``{{ChatGPT}} for
  {{Vulnerability Detection}}, {{Classification}}, and {{Repair}}: {{How Far
  Are We}}?,'' in {\em 2023 30th {{Asia-Pacific Software Engineering
  Conference}} ({{APSEC}})}, pp.~632--636, Dec. 2023.

\bibitem{khareUnderstandingEffectivenessLarge2024}
A.~Khare, S.~Dutta, Z.~Li, A.~{Solko-Breslin}, R.~Alur, and M.~Naik,
  ``Understanding the {{Effectiveness}} of {{Large Language Models}} in
  {{Detecting Security Vulnerabilities}},'' Oct. 2024.

\bibitem{tambergHarnessingLargeLanguage2024}
K.~Tamberg and H.~Bahsi, ``Harnessing {{Large Language Models}} for {{Software
  Vulnerability Detection}}: {{A Comprehensive Benchmarking Study}},'' May
  2024.

\bibitem{islamLLMPoweredCodeVulnerability2024}
N.~T. Islam, J.~Khoury, A.~Seong, M.~B. Karkevandi, G.~D. L.~T. Parra,
  E.~{Bou-Harb}, and P.~Najafirad, ``{{LLM-Powered Code Vulnerability Repair}}
  with {{Reinforcement Learning}} and {{Semantic Reward}},'' Feb. 2024.

\bibitem{zhouMultiLLMCollaborationDataCentric2024}
X.~Zhou, K.~Kim, B.~Xu, D.~Han, and D.~Lo, ``Multi-{{LLM Collaboration}} +
  {{Data-Centric Innovation}} = 2x {{Better Vulnerability Repair}},'' Mar.
  2024.

\bibitem{sloane-andersonBenchmarkingLargeLanguage2024}
R.~{Sloane-Anderson} and H.~Price, ``Benchmarking {{Large Language Models}} for
  {{Safe Software Development Advice}},'' June 2024.

\bibitem{espinhagasibaImSorryDave2023}
T.~Espinha~Gasiba, K.~Oguzhan, I.~Kessba, U.~Lechner, and
  M.~{Pinto-Albuquerque}, ``I'm {{Sorry Dave}}, {{I}}'m {{Afraid I Can}}'t
  {{Fix Your Code}}: {{On ChatGPT}}, {{CyberSecurity}}, and {{Secure
  Coding}},'' in {\em 4th {{International Computer Programming Education
  Conference}} ({{ICPEC}} 2023)} (R.~A. {Peixoto de Queir{\'o}s} and M.~P.
  Teixeira~Pinto, eds.), vol.~112 of {\em Open {{Access Series}} in
  {{Informatics}} ({{OASIcs}})}, (Dagstuhl, Germany), pp.~2:1--2:12, Schloss
  Dagstuhl -- Leibniz-Zentrum f{\"u}r Informatik, 2023.

\bibitem{heLargeLanguageModels2023}
J.~He and M.~Vechev, ``Large {{Language Models}} for {{Code}}: {{Security
  Hardening}} and {{Adversarial Testing}},'' in {\em Proceedings of the 2023
  {{ACM SIGSAC Conference}} on {{Computer}} and {{Communications Security}}},
  pp.~1865--1879, Nov. 2023.

\bibitem{bsimm2018}
G.~McGraw, S.~Migues, and B.~Chess, ``{Building Security in Maturity Model
  (BSIMM)},'' 2018.
\newblock Version 9.

\bibitem{microsoftSDL}
{Microsoft}, ``{Microsoft Security Development Lifecycle (SDL)}.''
  \url{https://www.microsoft.com/en-us/sdl}, Nov. 2018.
\newblock Accessed: 2025-01-13.

\bibitem{seiCertCoding}
R.~C. Seacord, ``{SEI CERT Secure Coding Standards}.''
  \url{https://wiki.sei.cmu.edu/confluence/display/seccode}, Dec. 2006.
\newblock Accessed: 2025-01-13.

\bibitem{seiCertTop10}
{Software Engineering Institute}, ``{SEI CERT Top 10 Secure Coding
  Practices}.''
  \url{https://wiki.sei.cmu.edu/confluence/display/seccode/Top+10+Secure+Coding+Practices},
  2018.
\newblock Accessed: 2025-01-13.

\bibitem{owasp}
{OWASP Foundation}, ``{The Open Web Application Security Project (OWASP)}.''
  \url{https://owasp.org/}, 2025.
\newblock Accessed: 2025-01-13.

\bibitem{mcdm}
{ScienceDirect}, ``{Multi-Criteria Decision Making - ScienceDirect Topics}.''
  \url{https://www.sciencedirect.com/topics/engineering/multi-criteria-decision-making},
  2025.
\newblock Accessed: 2025-01-13.

\bibitem{jiangTrustworthinessEvaluationMethod2014}
R.~Jiang, ``{A Trustworthiness Evaluation Method for Software Architectures
  Based on the Principle of Maximum Entropy (POME) and the Grey Decision-Making
  Method (GDMM)},'' {\em Entropy}, vol.~16, no.~9, pp.~4818--4838, 2014.

\bibitem{macekModelEvaluationCritical2021}
D.~Maček, I.~Magdalenić, and N.~Begičević~Ređep, ``A model for the
  evaluation of critical it systems using multicriteria decision-making with
  elements for risk assessment,'' {\em Mathematics}, vol.~9, no.~9, p.~1045,
  2021.

\bibitem{ningHybridMCDMApproach2020}
L.~Ning, Y.~Ali, and H.~Zhao, ``A hybrid mcdm approach of selecting lightweight
  cryptographic cipher base on iso and nist lightweight cryptography security
  requirements,'' {\em IEEE Access}, vol.~8, pp.~193695--193711, 2020.

\bibitem{dujmovicModelingAggregationSecurity2012}
J.~J. Dujmovi{\'c}, ``{Strong and Weak Simultaneity and Replaceability},'' {\em
  International Journal of Uncertainty, Fuzziness and Knowledge-Based Systems},
  vol.~20, no.~Supp01, pp.~1--29, 2012.
\newblock Foundational paper defining the logic operators for simultaneity and
  replaceability used in LSP.

\bibitem{dassoWebApplicationsSecurity2020}
A.~Dasso and A.~Funes, ``Web {{Applications Security Testing Evaluation}},'' in
  {\em 49 JAIIO - SIIIO (Simposio Argentino de Informática Industrial e
  Investigación Operativa)}, (San Luis, Argentina), pp.~102--114, SADIO, 2020.

\bibitem{weiChainofThoughtPromptingElicits2022}
J.~Wei, X.~Wang, D.~Schuurmans, M.~Bosma, B.~Ichter, F.~Xia, E.~Chi, Q.~Le, and
  D.~Zhou, ``Chain-of-thought prompting elicits reasoning in large language
  models,'' {\em Advances in Neural Information Processing Systems}, vol.~35,
  pp.~24824--24837, 2022.

\bibitem{juliet-test-suite}
P.~Arteau, ``Juliet test suite dataset.''
  \url{https://github.com/find-sec-bugs/juliet-test-suite/tree/master}, 2017.

\bibitem{wsvd-bench}
N.~Antunes, ``Wsvd bench dataset.''
  \url{https://github.com/nmsa/wsvd-bench/tree/master}, 2019.

\bibitem{owaspSQLInjectionPrevention2024}
{OWASP Foundation}, ``{SQL Injection Prevention Cheat Sheet},'' 2024.
\newblock Accessed: 2024-05-20.

\bibitem{cwe}
{MITRE Corporation}, ``Common weakness enumeration (cwe),'' 2025.
\newblock Accessed: 2025-01-13.

\bibitem{joernDocs}
{ShiftLeft Inc.}, ``Joern documentation.'' \url{https://docs.joern.io/}, 2025.
\newblock Accessed: 2025-08-26.

\end{thebibliography}

\end{document}